\documentclass[preprint]{elsarticle}
\usepackage{lineno,hyperref}
\usepackage{bm}
\usepackage{amssymb}
\usepackage{amsmath}
\usepackage{xcolor}
\usepackage{placeins}
\usepackage{caption}
\usepackage{subcaption}
\modulolinenumbers[5]

 \newcommand{\blue}[1]{\textcolor{black}{#1}}

\journal{arXiv}

\biboptions{authoryear}

\begin{document}


\begin{frontmatter}

\title{Vine Copula based portfolio level conditional risk measure forecasting}

\author[1]{Emanuel Sommer}
\author[2,3]{Karoline Bax}
\author[1]{Claudia Czado}

\address[1]{Department of Mathematics, Technical University of Munich, Munich, Germany}

\address[2]{Department of Economics and Management, University of Trento, Trento, Italy}
\address[3]{TUM School of Management, TUM Campus Heilbronn, Technical University of Munich, Heilbronn, Germany}

\begin{abstract}
\blue{
Accurately estimating risk measures for financial portfolios is critical for both financial institutions and regulators. However, many existing models operate at the aggregate portfolio level and thus fail to capture the complex cross-dependencies between portfolio components. To address this, a new approach is presented that uses vine copulas in combination with univariate ARMA-GARCH models for marginal modelling to compute conditional portfolio-level risk measure estimates by simulating portfolio-level forecasts conditioned on a stress factor. A quantile-based approach is then presented to observe the behaviour of risk measures given a particular state of the conditioning asset(s). In a case study of Spanish equities with different stress factors, the results show that the portfolio is quite robust to a sharp downturn in the American market. At the same time, there is no evidence of this behaviour with respect to the European market.}
The novel algorithms presented are ready for use through the R  package \textit{portvine}, which is publicly available on CRAN.
\end{abstract}


\end{frontmatter}

\section{Introduction}
\blue{Financial institutions and regulators rely on accurate risk measures to value financial instruments and portfolios. The current situation with the Covid-19 pandemic highlights the importance of accounting for market risk, which can manifest itself, for example, as high volatility clusters. The recommended risk measure for accurate estimation is the Expected Shortfall (ES), which improves upon the basic Value at Risk (VaR) by overcoming its theoretical shortcomings \citep{embrecht_1997,Artzner1999CoherentMO,ESalternativeVaR}.} However, many existing methods for estimating these risk measures lack the ability to adequately incorporate the \blue{potentially high dimensional dependency structure within financial portfolios and therefore fail to provide an accurate picture of reality.  In addition, practitioners often rely on sensitivity analyses of the risk measure at hand to further validate its robustness, as discussed in \cite{PESENTI2019654} and the references therein. \cite{PESENTI2019654} puts the portfolio manager's problem nicely in simple terms: "What does it take to break the model?". Following this, it is necessary to implement a one-day-ahead forecast of these risk measures in a portfolio setting under conditioning, which then allows for different stress testing approaches.}

\blue{To address the issue of modelling dependence between the components of the potentially high-dimensional portfolio, we chose to use vine copulas. More specifically, by incorporating marginal modelling and different copula specifications, we are able to accurately model high dimensional dependence. Since the introduction of vine copulas, many researchers have combined vine copula models with VaR risk measures. These include, but are not limited to, the work of \cite{berg2009models}, who focus on a four-component portfolio with marginal GARCH(1,1) modelling, but with no rolling window, no forecasts and one-day ahead VAR, and \cite{chollete2009modeling}, who use canonical vine autoregressive models with a larger number of S\&P 500 stocks across 10 different sectors. While they include one-day ahead forecasts, they do not consider a rolling window. Finally, \cite{brechmann13} develop a regular Vine copula-based factor model for asset returns and compare it with the popular DCC models.} Recently, some scholars have also attempted to overcome the challenge of correctly representing portfolio dependence. \blue{The approach in \cite{cryptovar} is almost identical to that of \cite{mohamed}, but differs in the use of robust GARCH models instead of ARMA-GARCH models. It successfully applies the approach to portfolios of cryptocurrencies and also covers the ES.} \blue{A similar approach using rolling windows and marginal ARMA$(1,1)$-GARCH$(1,1)$ models together with a vine copula has already been proposed by \cite{nagler2019model}. } \cite{monika} and \cite{mutualinfo} both propose the same unconditional risk measure estimation approach as \cite{mohamed}, combining univariate ARMA-GARCH models with regular vine copula models as the flexible dependence model, followed by a Monte Carlo approach to estimate the risk measures, but all lack ES coverage and do not provide software or optimised algorithms for effective estimation.  \blue{Also, most of them do not take into account the problem of non-stationarity of the time series, as they do not include a rolling window approach. Our work overcomes these limitations. }

\blue{Furthermore, it is generally accepted that stress testing with a single stressor is useful for sensitivity analysis. Previous approaches without copula tools include a combination of the multivariate joint Gaussian distribution and the use of the conditional distribution of assets given the stressor variable \citep{dempster2002risk}. When considering copula-based methods, the work of \cite{brechmannhendrich2013} uses C-vines and focuses on overall systemic risk analysis. The lack of sensitivity analysis in the area of risk measures is also addressed in the work of \cite{PESENTI2019654}. 
They propose a model-independent reverse sensitivity testing framework, which they argue is best suited for contexts where the model for calculating the risk measures is too complex to derive a model-based sensitivity approach. However, for the presented risk estimation procedure proposed in this paper, we are able to leverage the setup of the approach and derive a model-based sensitivity analysis.}

\blue{As highlighted along the way, these works do not provide a detailed and optimised algorithm for the rolling window approach and a packaged, easy-to-use implementation, nor do they address the problem of assessing the robustness of the estimated risk measures, e.g. through stress testing. In our work, we overcome these limitations and provide a multifaceted contribution.}

 \blue{First, we provide the details of the missing algorithms for the optimised rolling window estimation. Then, our main contribution is to extend the previous unconditional approach by introducing a conditional element. More specifically, we effectively measure the risk of \blue{potentially} large asset portfolios in a conditional setting using a stress factor. In order to be able to model large dependence structures between the components of the portfolio, we use the popular class of vine copulas, more specifically D-vine copulas (\cite{joe2014dependence}, \cite{kraus17}, \cite{Czado2019}). This allows us to analytically determine the conditional distribution of the components given any stress factor we choose.  Then, by simulating the portfolio-level return forecasts conditional on the stress factor, we are able to estimate the conditional portfolio-level risk estimates. We include both VaR and the superior risk measure ES. By considering different quantiles of the stressors, we can gain valuable insights in different stress testing scenarios. This allows for sensitivity analysis of the risk measures along the conditional asset of choice. 
In addition, as the backbone of our simulation-based approach, we present a novel algorithm for sampling from the stressors of a D-vine copula, implemented in the published software package \textit{portvine} (\cite{portvine}) for the statistical programming language R\footnote{the package can be found at https://CRAN.R-project.org/package=portvine}.  The package is capable of estimating the unconditional, an optimised version of the approach proposed by \cite{mohamed}, as well as the conditional risk measures as proposed in this paper. Finally, we include a \blue{medium sized} case study on Spanish stocks to show the practical application of the proposed methods. }

The paper is structured as follows.  Section 2 defines VaR and ES, while Section 3 introduces models for portfolios subject to a stress factor.  Section 4 then focuses on risk measure prediction \blue{and rolling window setup including detailed information on the algorithm and software performance details}.  Section 5 shows the practical application in a case study \blue{and the results from a concept simulation.} Finally, Section 6 concludes the paper.

\section{Risk Measures}

Most financial analyses still rely on the VaR which was initially introduced by JP Morgan as pointed out by \citet{pfaff2016financial}.  The VaR  of the random variable $R^P_t$ representing the value of an asset or portfolio P at time $t$ is denoted by $\bm{VaR}_{\alpha}^{P,t}$ and defined as  
\begin{equation}
  \label{eq:var}
  \bm{VaR}_{\alpha}^{P,t} := sup\{r|F_{R^P_t}(r) \leq \alpha\} = Q_{R^P_t}(\alpha),
\end{equation}
where $F_{R^P_t}$ and $Q_{R^P_t}$ are the distribution and quantile function of $R^P_t$, respectively.
It can be interpreted as the value that $R^P_t$ will only fall below with a probability of $\alpha$. 

Assuming that an  i.i.d. sample $r^{P, s}_t$, for $s \in\{1,\dots,S\}$ is available from the random variable $R^P_t$ the $\bm{VaR}_{\alpha}^{P,t}$ is estimated by its emprirical quantile
\begin{equation}
    \label{eq:varest}
    \widehat{\bm{VaR}}^{P,t}_\alpha = \widehat{\bm{VaR}}(\{r^{P, s}_t|s \in\{1,\dots,S\}\}, \alpha) = \hat{Q}_{\{r^{P, s}_t|s \in\{1,\dots,S\}\}}(\alpha).
\end{equation}
Here $\hat{Q}_{Set}(\alpha)$ denotes the standard empirical quantile function based on a $Set$ of samples evaluated at the confidence level $\alpha$. 
However it is well known that the VaR has shortcomings. It is
not a coherent risk measure as it is not  subadditive as shown in \cite{Artzner1999CoherentMO} and \cite{ESalternativeVaR}. This major weakness  makes it hard to assess overall risk when managing multiple portfolios.
Further, this  measure does not  contain any information about the value of the asset or portfolio if its value falls below the VaR at the chosen confidence level. It only gives an upper bound for this value, therefore it is more informative to consider the expected shortfall (\textbf{ES}) with confidence level $\alpha \in (0,1)$ defined as
\begin{equation}
  \label{eq:es}
  \bm{ES}_{\alpha}^{P,t} = \mathbb{E}[R^P_t | R^P_t \leq \bm{VaR}_{\alpha}^{P,t}] =  \frac{1}{\alpha} \int_{0}^{\alpha} \bm{VaR}_{u}^{P,t} du.
\end{equation}
It can be interpreted as the average value of $R^P_t$ if the asset or portfolio drops in value below the $\bm{VaR}_{\alpha}^{P,t}$. 

Contrary to the VaR,  the ES is indeed a coherent risk measure which was proven for example in \cite{Tasche2002ExpectedSA} and thus has the desired subadditivity property.  So indeed the ES can mitigate the major shortcomings of the VaR but for example the law invariance weakness still applies to the ES.  In the literature the expected shortfall is often called the conditional Value at Risk (\textbf{CVaR}) and the representation using the expected value for general also non absolutely continuous random variables is often called the tail conditional expectation (\textbf{TCE}) which is equivalent to the ES in the continuous case.

Assuming again the availability of the sample $r^{P, s}_t$, for $s \in\{1,\dots,S\}$, we can estimate 
$ \bm{ES}_{\alpha}^{P,t}$ by the empirical mean over all samples with $r^{P, s}_t \leq  \bm{VaR}^{P,t}_\alpha$.
    
\section{Vine copula models for portfolio constituents given a stress factor level} \label{VineCop}

To allow for more general dependence structures among the portfolio components we want to utilize the popular model class of regular vine copulas \citep{joerecursion, bedford2001probability,graphsbedoford,aas2009pair}.
In particular we are interested in the sub class of D-vine copulas, which allows us to determine analytically (without integration) the conditional distribution of the components given a stress factor. For the convenience of the reader we recall the most basic ingredients for a D-vine copula. 
\subsection{D-vine basics}
The copula approach based on \citet{sklar1959fonctions} allows to separate the dependence from the marginal distributions, since any distribution function $F$ of a $d$ dimensional random vector $\bm X$ with marginal distributions $F_j$ and marginal density functions $f_j$ for $j=1,\ldots, d$ can be represented as
\begin{equation}
\label{sklar-cdf}
F(x_1,...,x_d) = C(F_1(x_1),...,F_d(x_d)).
\end{equation}
Here $C$ denotes the $d$ dimensional copula, which is the distribution function of a random vector $\bm U$ taking only values in $[0,1]^d$ and has uniform margins. The copula in \eqref{sklar-cdf} is uniquely defined for absolutely continuous distributions, which we assume in the following. The associated density can then also expressed as 
 $f(x_1,...,x_d) = c(F_1(x_1),...,F_d(x_d)) f_1(x_1)\cdots  f_d(x_d)$,
 where $c$ is the corresponding copula density.
 By inversion of Equation (\ref{sklar-cdf}) we can use any $d$-dimensional distribution function to obtain the corresponding copula. Examples 
are the multivariate Gaussian and the Student $t$ copula. Multivariate
 Archimedean copulas are another class of parametric copulas that are built directly using generator functions. The Gumbel, Clayton and Frank copula families are prime examples. 
Bivariate or pair copulas ($d=2$) are the building blocks
for vine copulas in $d$ dimensions. Conditioning (\cite{joerecursion}, \cite{graphsbedoford}) is required to glue together the pair copulas and thus we need the conditional 
 density $f_{1|2}$ and distribution function $F_{1|2}$ which are given by
{\small
\begin{eqnarray}
f_{1|2}(x_1|x_2) & = & c_{1 2 }(F_1(x_1), F_2(x_2)) f_1(x_1), \label{cond-den}\\
F_{1|2}(x_1|x_2) 
 & = &\frac{\partial}{\partial F_2(x_2)}
 C_{1 2}(F_1(x_1),F_2(x_2)) =  \frac{\partial}{\partial v}
 C_{1 2}(F_1(x_1),v)|_{v= F_2(x_2)}\label{cond-dist}.
\end{eqnarray}
}
We show the construction for $d=3$ using
the  factorization 
\begin{equation}
\label{fac3}
f(x_1,x_2,x_3)=f_{3|12}(x_3|x_1,x_2) f_{2|1}(x_2|x_1)f_1(x_1)
\end{equation}
and study each part separately. Here $F_{j|D}$ and $f_{j|D}$  denote the conditional distribution or density function of $X_j$ given  $\bm X_{D}$, respectively.
To determine $f_{3|12}(x_3|x_1,x_2)$ we consider the bivariate conditional density $f_{13|2}(x_1,x_3|x_2)$. 
The copula density  $c_{13;2}(\cdot,\cdot;x_2)$ denotes the copula density associated with the conditional distribution of $(X_1,X_3)$ given $X_2=x_2$. Using  
 \eqref{sklar-cdf} in terms of densities for $f_{13|2}(x_1,x_3|x_2)$ gives
\begin{align}
\label{fac3.1-bi}
    f_{13|2}(x_1,x_3|x_2) & =  c_{13;2}(F_{1|2}(x_1|x_2),F_{3|2}(x_3|x_2);x_2) f_{1|2}(x_1|x_2)
    f_{3|2}(x_3|x_2).
    \end{align}
The density of $X_3$ given $X_1=x_1,X_2=x_2$, denoted by $f_{3|12}(x_3|x_1,x_2)$, is determined from \eqref{cond-den}
applied to \eqref{fac3.1-bi}, yielding
\begin{equation}
\label{fac3.1}
f_{3|12}(x_3|x_1,x_2) =
c_{13;2}(F_{1|2}(x_1|x_2),F_{3|2}(x_3|x_2);x_2) f_{3|2}(x_3|x_2).
\end{equation}
Finally, direct application of  \eqref{cond-dist}
gives 
\begin{eqnarray}
\label{fac3.2}
f_{2|1}(x_2|x_1) & = &c_{12}(F_1(x_1),F_2(x_2)) f_2(x_2),\\
\label{fac3.3}
f_{3|2}(x_3|x_2) & = &c_{23}(F_2(x_2),F_3(x_3))f_3(x_3).
\end{eqnarray}
Inserting  \eqref{fac3.1}--\eqref{fac3.3} into  \eqref{fac3} yields a pair copula decomposition
of an arbitrary three dimensional
density $f(x_1,x_2,x_3)$  as
\begin{align}
\label{pcc3}
f(x_1,x_2,x_3)  & =  c_{13;2}(F_{1|2}(x_1|x_2),F_{3|2}(x_3|x_2);x_2)
\times c_{23}(F_2(x_2),F_3(x_3))\\
& \times  c_{12}(F_1(x_1),F_2(x_2)) f_3(x_3) f_2(x_2) f_1(x_1).
\nonumber
\end{align}
Two further decompositions are available when one changes the conditioning order in \eqref{fac3}.
Often the simplifying assumption is made to facilitate easy estimation. It assumes that the conditional pair copula  in \eqref{pcc3}  for
$c_{13;2}$ does not depend on the conditioning value $x_2$, the dependence on $c_{13;2}$ on $x_2$ is solely captured by the arguments $F_{1|2}(x_1|x_2)$ and $F_{3|2}(x_3|x_2)$. So we assume that 
$c_{13;2}(\cdot, \cdot;x_2)=c_{13;2}(\cdot, \cdot)$ holds for $x_2$. This turns the decomposition \eqref{pcc3} into a construction approach for trivariate densities. A trivariate copula density $c$ arises, when we use as marginal distributions the uniform distribution in \eqref{pcc3}.

This approach to construct densities can be extended to $d$ dimensions. For this a framework is needed to identify the conditioning order, which was developed in \citet{graphsbedoford}. It requires the specification of the vine tree structure. This graphical structure is build from $d-1$ linked trees, the first one has all $d$ variables as nodes, while subsequent trees are formed by having nodes being the edges of the previous trees and edges are allowed by the so called proximity condition. This means that two nodes in the current tree can only be connected when they (considered as edge in the previous tree) share a node in the previous tree. For a precise definition and illustrations see for example Chapter 5 of \citet{Czado2019}.



\begin{figure}
    \centering
    \includegraphics[width = 0.6\textwidth]{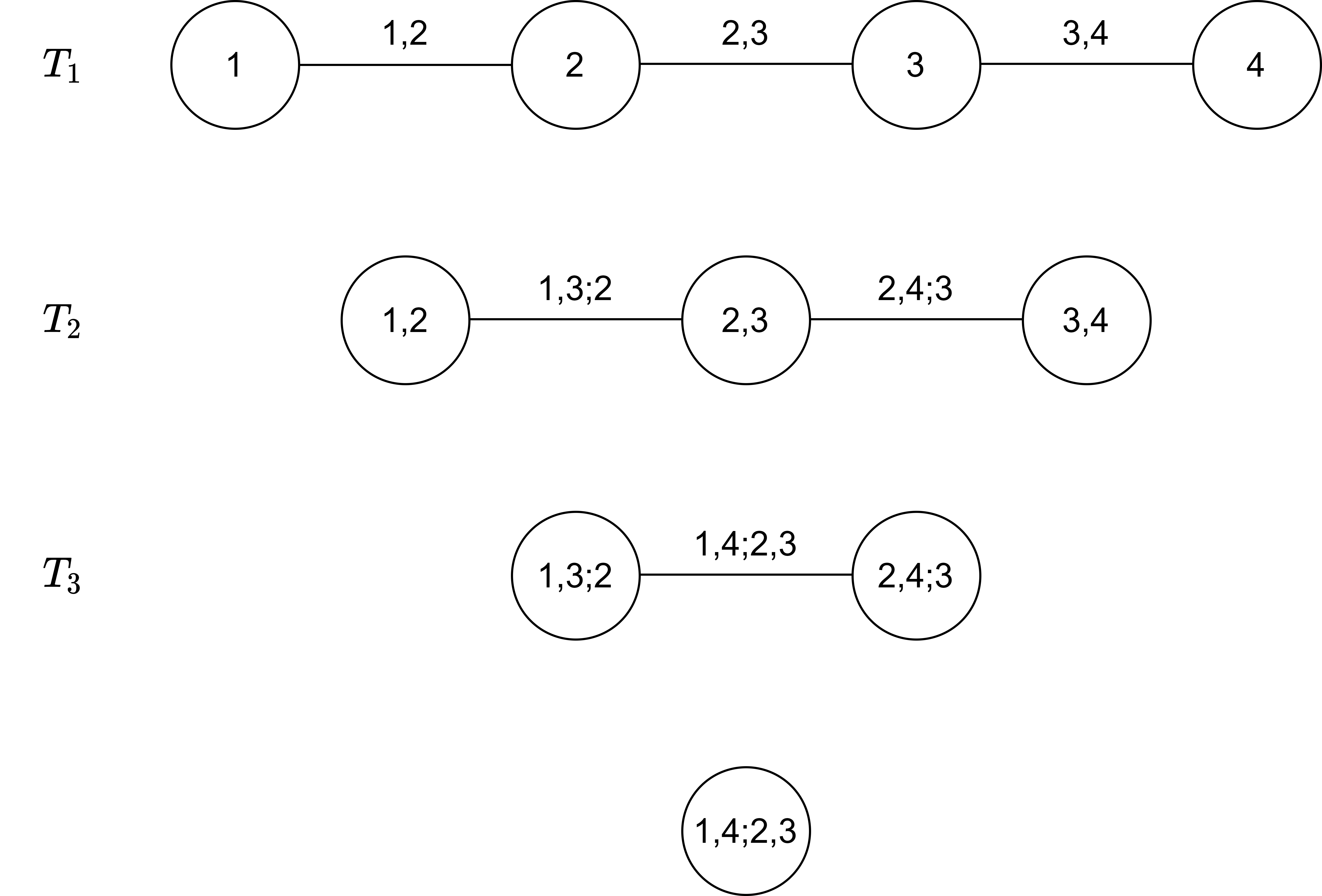}
    \caption{Exemplary D-vine tree sequence on 4 elements.}
    \label{fig:dvine4}
\end{figure}

One simple class of vine tree structures is the D-vine structure, where all trees have a path structure. An  example 
for $d=4$ is displayed in Figure \ref{fig:dvine4}. One property of a D-vine tree structure is that once the first tree is specified, the remaining trees are determined uniquely by the proximity condition. So a D-vine tree structure is specified by the order of the variables in the first tree. Further the nodes at the end of each tree are called the leaf nodes.

To build up the joint density all edges in the vine tree structure are associated with (conditional) pair copulas. Using  the abbreviation $i:j=(i,i+1,\ldots,j)$ for integers $i<j$ we can express the joint density of a D-vine distribution as
  \begin{equation}
  \label{dvine}
f_{1,\ldots,d}(x_1,\ldots,x_d)
 = \bigg [\prod_{j=1}^{d-1} \prod_{i=1}^{d-j}
c_{i,(i+j);(i+1)\ldots,(i+j-1)} \bigg ] \cdot \bigg[\prod_{k=1}^d
f_k(x_k) \bigg],
\end{equation}
where  $C_{j|D}$ and $c_{j|D}$  denote the conditional distribution  or density function of $U_j$ given  $\bm U_{D}$, respectively for $(U_1,\ldots,U_d)$ has copula density $c$. Additionally we used the abbreviation $c_{i,j|D}=c_{i,j|D}(F_{i|D}(x_i|\bm x_D),F_{j|D}(x_i|\bm x_D))$, where $\bm x_D= \{x_k, k\in D\}$.
The D-vine copula density to Figure \ref{fig:dvine4} is therefore  given by
 \begin{align}
\label{eq:dvine4ex}
    c_{1234}&(u_1,u_2,u_3,u_4) = c_{12}(u_1,u_2) \times c_{23}(u_2,u_3) \times c_{34}(u_3,u_4)\nonumber\\ \nonumber
    &\times c_{13;2}(C_{1|2}(u_1|u_2),C_{3|2}(u_3|u_2)) \times c_{24;3}(C_{2|3}(u_2|u_3),C_{4|3}(u_4|u_3))\\
    &\times c_{14;23}(C_{1|23}(u_1|u_2, u_3),C_{4|23}(u_4|u_2, u_3)).
\end{align}
In general 
\cite{joerecursion} showed a recursion for the conditional distribution function $C_{j|D}$ with the conditioning index set $D$ satisfying  $i \in D$ and $D_{-i} = D \setminus \{i\}$. For this we first define the $h$ functions associated with any pair copula $C_{ij;D}$ (independent of $\bm u_D$ by the simplifying assumption) as 
\begin{equation}
    h_{i|j;D}= \frac{\partial}{\partial u_2} C_{ij;D}(u_1,u_2) \mbox{ and }
    h_{j|i;D}= \frac{\partial}{\partial u_1} C_{ij;D}(u_1,u_2). 
\end{equation}
This allow us to express $C_{j|D}(u_j|\bm{u}_D)$ as
\begin{equation}
\label{cdfrecurs}
    C_{j|D}(u_j|\bm{u}_D) = h_{j|i;D_{-i}}\bigg(C_{j|D_{-i}}(u_j|u_{D_{-i}})|C_{i|D_{-i}}(u_i|u_{D_{-i}})\bigg).
\end{equation} 
Only pair copulas occurring in lower trees are needed for the recursion \eqref{cdfrecurs}.

Estimation in vine copula based models in general follow a two step procedure. In the case of multivariate time series data $r_{tj},t=1,\ldots, T, j=1,\ldots, d$ we first fit appropriate univariate time series such as ARMA-GARCH models for each component $r_{tj},t=1,\ldots, T$ to remove the serial dependence. The resulting standardized residuals $z_{tj},t=1,\ldots,T$ are then an approximate i.i.d sample from the utilized innovation distribution $F_j$ for component j. However the vectors 
$(z_{t1},\ldots,z_{td})$ for $t=1,\ldots, d$ are independent realizations from a multivariate distribution. In the next step we transform the standardized residuals $z_{tj}$ to the copula scale $u_{tj}=F_j(z_{tj})$ and use this multivariate sample to estimate an appropriate vine copula model. For the estimation of the parameters of a vine copula model the fast sequential estimation method as discussed for example in Chapter 7 of \cite{Czado2019} can be applied.

\subsection{Selecting the D-vine order for conditional portfolio risk measure estimation}\label{DvineSetup}

Since we want to estimate conditional portfolio risk measures based on $d$ returns of assets $A_1, \dots, A_d$ conditional on a stress level of the market index $I$, we want to find the D-vine copula model with the index $I$ as the last node in Tree 1 and the assets are ordered by an appropriate dependence strength.


 Recall that the ordering of the nodes in the first tree of an  D-vine tree  structure is sufficient to specify the complete vine tree structure. So we look for an ordering given by the permutation $j_1,\dots,j_d$ of the indices $1,\dots,d$ of the nodes in the first tree. Therefore the first tree is  specified as
\begin{equation}
    \label{eq:Dordering}
    A_{j_d} - A_{j_{d-1}} - \dots - A_{j_1} - I.
\end{equation}

We introduce now a further scale on which we measure the dependence strength. First the scale of the standardized residuals $z_{tj}$ is inappropriate, since it includes the effect of the marginal distributions $F_j$. The copula data $u_{tj}$ has the marginal effects removed but lives on $[0,1]$ and thus no easy estimation of conditional dependence measures are available. Thus we make a further transformation to normalized scores defined by $N_{tj}=\Phi^{-1}(u_{tj})$, where $\Phi$ is the distribution function of a standard normal distribution $N(0,1)$. Then $N_{tj}, t=1,\ldots, T$ is an approximate i.i.d sample of $N(0,1)$. For determining the order we are most concerned with central dependence measures, thus we use empirical estimates of the Pearson correlation (for the first tree) and partial correlations (higher order trees) using the normalized scores. Here we ignore the influence of a possibly non Gaussian copula present among the normalized scores. This follows since the multivariate normal density occurs in \eqref{dvine}, when all pair copulas are set to the bivariate Gaussian copulas with parameters set to the associated partial correlations.

\begin{figure}
    \centering
    \includegraphics[width = 0.7\textwidth]{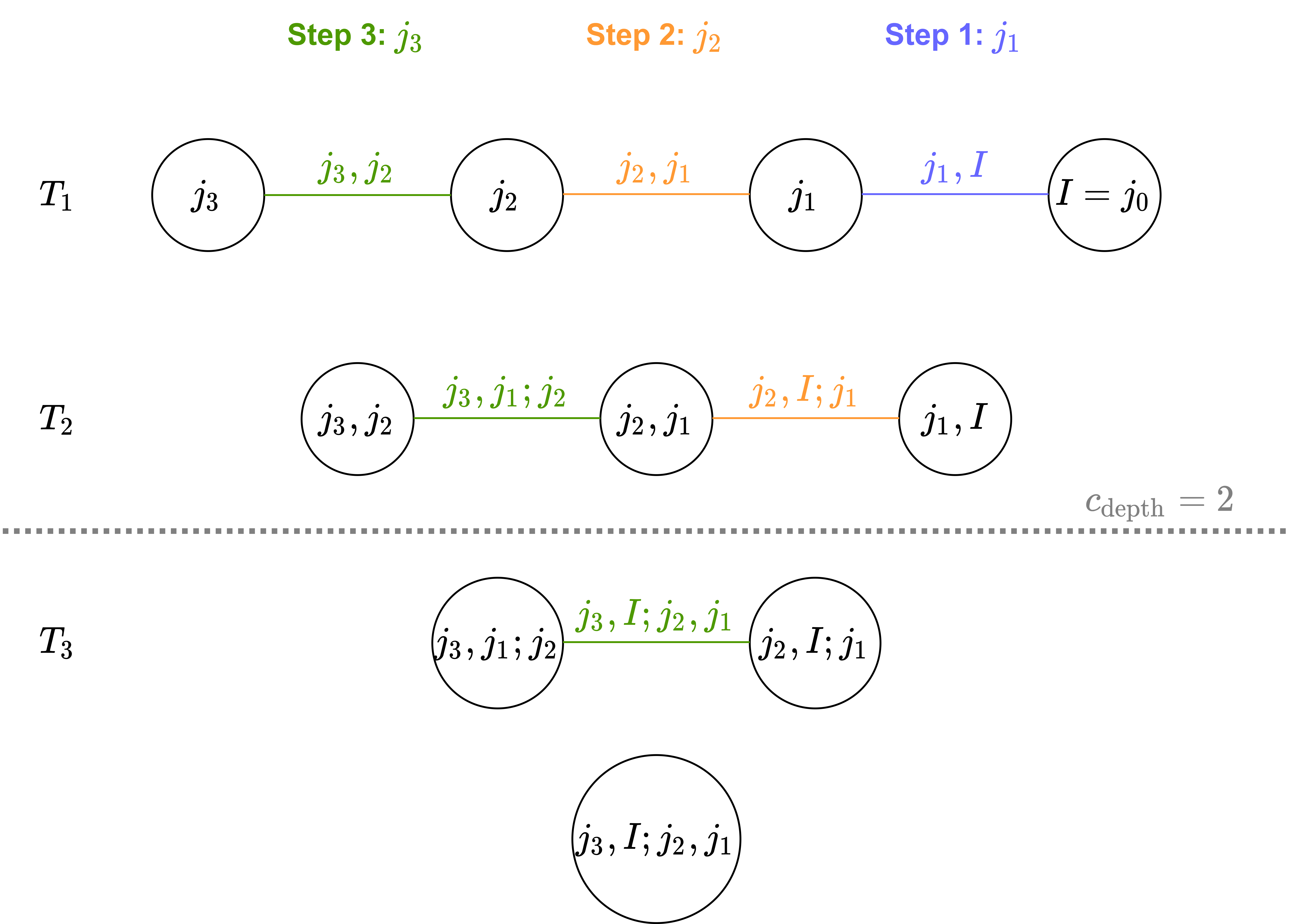}
    \caption{Illustration for the D-vine ordering algorithm in 4 dimensions.}
    \label{fig:conddvine4colored}
\end{figure}

We now describe our ordering algorithm in general and illustrate it for  $d=4$, i.e  we consider the 4 dimensional copula $c_{j_3j_2j_1I}$ density specified by the pair copula densities $c_{j_1,I}, c_{j_2,j_1}, c_{j_3,j_2}, c_{j_2,I;j_1}, c_{j_3,j_1;j_2}$ and $c_{j_3,I;j_2,j_1}$. Figure \ref{fig:conddvine4colored} illustrates this D-vine copula, where each edge corresponds to one of these pair copula densities.  We determine iteratively the permutation $j_1$ up to $j_d$ by first determining dependence weights for each pair copula term and then maximizing the sum of these weights. For the pair copula term $c_{ij}$ we use the associated Pearson correlation estimate $\hat{\rho}_{ij}$ based on the normalized scores, while for $c_{ij;D}$ we use the empirical partial correlation
$\hat{\rho}_{ij|D}$
again based on the normalized scores. This does not require to construct copula data for Trees 2 on wards to get conditional Kendall's $\tau$ estimates to be used as alternative. In Step 1 we choose $j_1$ such that $\hat{\rho}_{jI}$ is maximal among $j=1,\ldots,d$. In the following step we search for $j_2$, which maximizes $\hat{\rho}_{j,j_1}+\hat{\rho}_{j,I|j_1}$ for $j\in 1:d\setminus j_1$. The following step then searches for $j_3$ and is illustrated in Figure \ref{fig:conddvine4colored} in green. In general, we determine the index $j_k$ such that the sum of all the edge weights corresponding to the additional edges that were not yet specified but contain only $j_k, j_{k-1},\dots,I$ is maximized. The needed edges in each step are highlighted in different colors in Figure \ref{fig:conddvine4colored}.

This greedy process does of course not guarantee a global optimum as it makes local choices.
Especially for high dimensions however this approach accounting for all edges can be further simplified by introducing the cutoff depth $c_{\text{depth}}$. This implies that only all edge weights up to the specified depth will be used for the calculation of the ordering. This cutoff value $c_{\text{depth}}=2$ is indicated in Figure \ref{fig:conddvine4colored} by a dotted line so all edges below the line are set to independence copulas and thus have zero weights.

\subsection{Conditional simulation}
After the ordering is determined according to the algorithm  given in Section 3.2, we fit the 
D-vine copula $C_{j_1,\dots,j_d,I}$ with the order specified in \eqref{eq:Dordering}, i.e select appropriate families and estimate associated parameters for each pair copula term.
To set a stress level for the market index $I$, which is independent of marginal parameters of the market time series model, we note that a high copula value such as $u_I=.95$, results in a large standardized residual $z_{T+1,I}=F^{-1}_I(.95)$ and thus to a high one day ahead forecast for $R_{T+1,I}$. Thus this way of proceeding we can choose the value $\alpha \in [0,1]$ for $u_I$ to assess the effect of different market stress levels. Once we have chosen such an stress level $u_I=\alpha$, then we want to be able to generate replicates for the conditional distribution of $U_{j_d},\ldots U_{j_1}$ given $U_I=\alpha$ using the fitted D-vine copula. 

For notational ease we now assume that the ordering in \eqref{eq:Dordering} is given by $A_1-A_2- \cdots A_d - I$. Next we  apply the transform introduced by \cite{rosentrans} to simulate from $C_{1,\ldots,d|I}(\cdot,\ldots,\cdot|U_i=\alpha)$. More precisely 
\begin{align}
\label{rosenblatt}
    &\text{Sample } w_j, \overset{i.i.d.}{\sim} \mathcal{U}(0,1), j = 1,\dots,d  \nonumber \\
    &u_d := C^{-1}_{d|I}(w_d|U_I=\alpha) \nonumber \\
    &u_{d-1} := C^{-1}_{d-1|d,I}(w_{d-1}|U_{d}=u_{d},U_I=\alpha) \\
    &\vdots \nonumber \\
    &u_1 := C^{-1}_{1|2,\ldots,d,I}(w_1|U_2=u_2, \ldots, U_d=u_{d},\dots, U_I=\alpha), \nonumber  
\end{align}
then $(u_1,\ldots, u_d)$ is a realization from  $C_{1,\ldots,d|I}(\cdot,\ldots,\cdot|U_i=\alpha)$. 

\begin{figure}
    \centering
    \includegraphics[width = 0.7\textwidth]{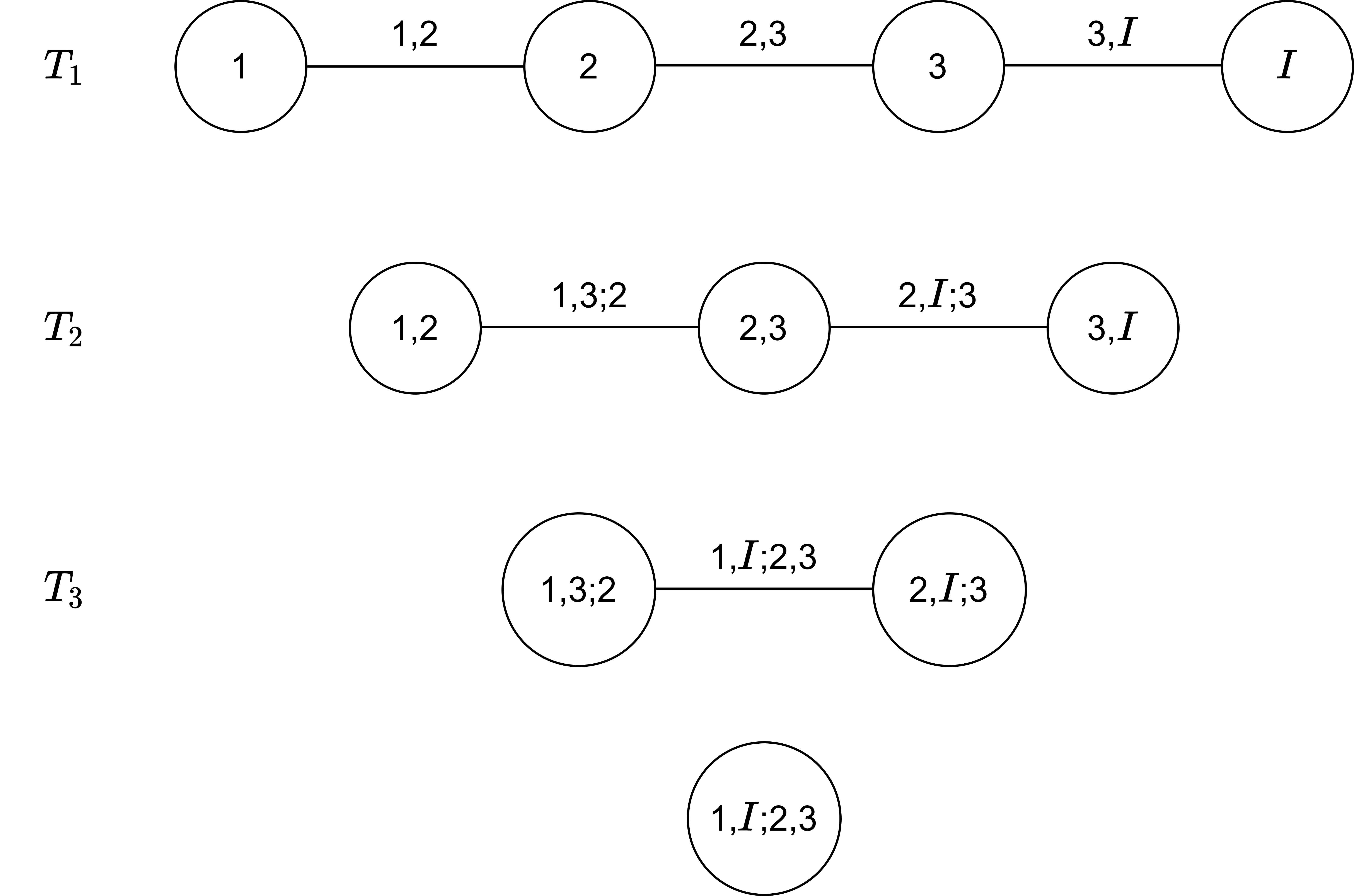}
    \caption{Illustration of the tree sequence of a 4-dimensional D-vine with the market index $I$ as the rightmost leaf node.}
    \label{fig:condvine4ex}
\end{figure}

We now illustrate for $d= 4$ how the required quantile functions in  \eqref{rosenblatt} are determined using appropriate (inverse) $h$ functions.
For this we start with a fitted D-vine copula $C_{1,2,3,I}$. So the following pair copulas were fitted: $C_{1,2},C_{2,3},C_{3,I}$, $C_{1,3;2},C_{2,I;3}$ and $C_{1,I;2,3}$. The corresponding D-vine tree structure  is displayed in Figure \ref{fig:condvine4ex}.
The required calculations are now shown in detail for $d=4$:
\begin{align*}
    &u_I=\alpha\\
    &w_1,w_2,w_3 \overset{i.i.d.}{\sim} \mathcal{U}(0,1)\\
    w_3 &= C_{3|I}(u_3|u_I) \\
    &= h_{3|I}(u_3|u_I) \\
    &\quad\Rightarrow u_3 = h_{3|I}^{-1}(w_3|u_I), \text{ so } u_3 \text{ is a sample from } C_{3|U_I=u_I}.\\
    w_2 &= C_{2|3,I}(u_2|u_3,u_I)\\
    &= \frac{\partial}{\partial C_{I|3}(u_I|u_3)} C_{2,I;3}(C_{2|3}(u_2|u_3),C_{I|3}(u_I|u_3))\\
    &= \frac{\partial}{\partial v_2} C_{2,I;3}(h_{2|3}(u_2|u_3),v_2) \mid_{v_2 = h_{I|3}(u_I|u_3)}\\
    &= h_{2|I;3}(h_{2|3}(u_2|u_3),h_{I|3}(u_I|u_3))\\
    &\quad\Rightarrow u_2 = h_{2|3}^{-1}(h_{2|I;3}^{-1}(w_2|h_{I|3}(u_I|u_3))|u_3), \text{ so } u_2 \text{ is a sample from } \\
    & C_{2|3 = u_3,U_I = u_I}.\\
     w_1 &= C_{1|2,3,I}(u_1|u_2,u_3,u_I)\\
    &= \frac{\partial}{\partial C_{I|2,3}(u_I|u_2,u_3)} C_{1,I;2,3}( C_{1|2,3}(u_1|u_2,u_3), C_{I|2,3}(u_I|u_2,u_3))\\
    &= h_{1|I;2,3}(C_{1|2,3}(u_1|u_2,u_3)| C_{I|2,3}(u_I|u_2,u_3))\\
    &= h_{1|I;2,3}(h_{1|3;2}(C_{1|2}(u_1|u_2)| C_{3|2}(u_3,u_2)) | C_{I|2,3}(u_I|u_2,u_3))\\
    &= h_{1|I;2,3}(h_{1|3;2}(h_{1|2}(u_1|u_2)| h_{3|2}(u_3,u_2)) | C_{I|2,3}(u_I|u_2,u_3))\\
    &\quad\Rightarrow u_1 = h_{1|2}^{-1}(h_{1|3;2}^{-1}(h_{1|I;2,3}^{-1}(w_1|C_{I|2,3}(u_I|u_2,u_3))| h_{3|2}(u_3,u_2))|u_2), \\
    &\quad\text{so with } C_{I|2,3}(u_I|u_2,u_3) = h_{I|2;3}(h_{I|3}(u_I|u_3)|h_{2|3}(u_2,u_3))\\
    &\quad  u_1 \text{ is a sample from } C_{1|2 = u_2,3 = u_3,I = u_I}.\\ 
\end{align*}

\noindent Finally $(u_1,u_2,u_3)$ is a realization of the conditional distribution of $(U_1,U_2,U_3)$ given $U_I = \alpha$ and it can be easily checked that all the needed bivariate copulas are estimated by the assumed D-vine. A formulation of the general conditional sampling algorithm can be found in \cite{SommerThesis}.


\section{Conditional portfolio risk measures forecasting}

\noindent In the following we are going through the risk measure forecasting steps.

\subsection{ES($P|u^I$), VaR($P|u^I$)}

We define a portfolio $\Omega = \{w_j,r_t^{A_j} | t = 1, \dots, T; j = 1, \dots, d\}$, where $r_t^{A_j}$ is the realized log return of the asset $A_j$ at time $t$. Here  the length of the fitting window for the marginal financial time series models is defined as  ${\Gamma} < T \in \mathbb{N}$. One assumes that this is also the time index of the most recent asset return observation when starting the one step ahead or rolling window approach. So for the first marginal time series model one uses all the available historic data. In Figure \ref{fig:onestep_rm_estimate_params} the associated window to $\Gamma$ is highlighted in blue.  Additionally,  ${S} \in \mathbb{N}$ is the number of simulated log returns for the risk measure estimation. The market index $I$  is observed over the same time period as the portfolio $\Omega$, the log returns are again denoted by $r_t^{I}$.  Then, let  $j_1, \dots, j_d$ be a to be determined permutation of the indices of the assets $1, \dots, d$. Lastly, $\alpha^I \in (0,1)$ is the confidence level of the estimated quantile from the marginal market index distribution. This confidence level is the conditioning value for the final risk measure estimate on the copula scale. How this fits into the estimation process can be found in detail below.

\begin{figure}[tbh!]
    \centering
    \includegraphics[width = \textwidth]{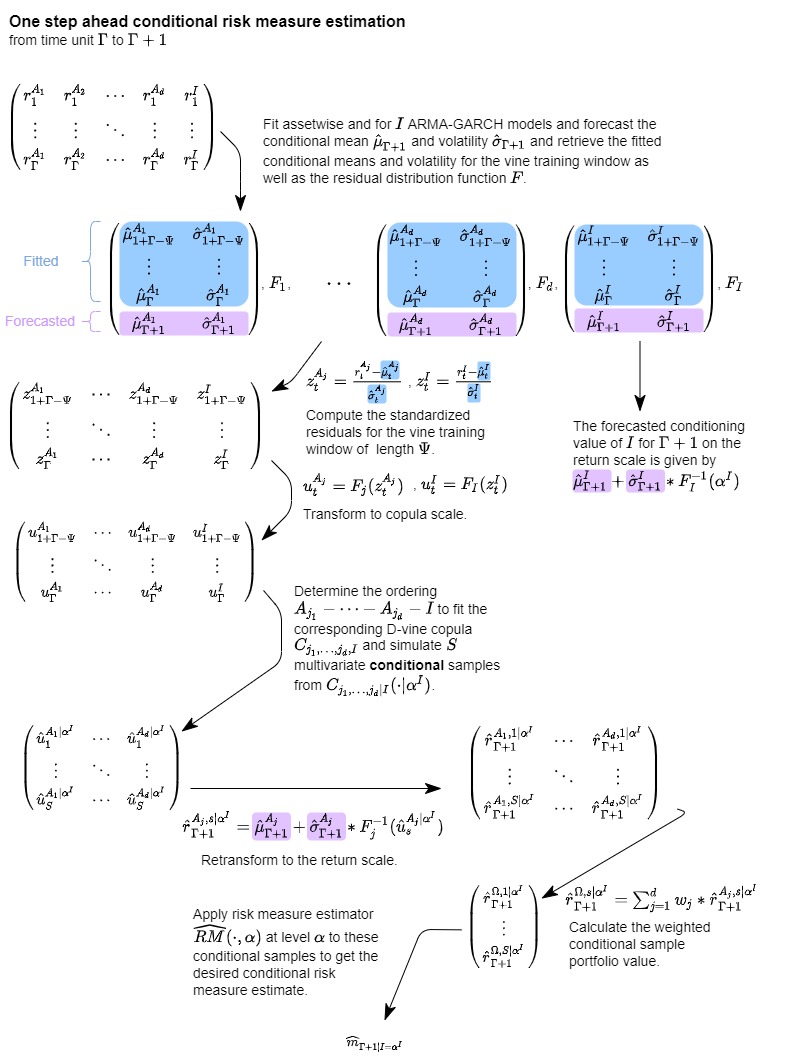}
    \caption{Illustration of the one step ahead single conditional risk measure estimation approach using ARMA-GARCH marginal and D-vine dependence models.}
    \label{fig:onestep_rm_estimate_cond_corrected}
\end{figure}

Figure \ref{fig:onestep_rm_estimate_cond_corrected} illustrates the flow of the data along the applied methods and transformations for the one step ahead single conditional risk measure estimation approach:

\begin{itemize}
    \item Starting with the log returns of the $d$ assets $r_t^{A_j}$, we fit an adequate ARMA-GARCH model for each of the assets log return series and to the series corresponding to the conditional asset $I$. An ARMA(1,1)-GARCH(1,1) is  often a good starting point as  mentioned in \cite{tsay2010}.
    \item Then, in order to model the dependence between the assets and the index $I$ we use the now assumed i.i.d. standardized residuals $z_t^{A_j}$ and $z_t^I$ to determine the copula data $u_t^{A_j}$ and $u_t^I$ applying the  marginal innovation distributions $F_j$ and $F_I$, respectively.
    \item The obtained copula scale data is then used to estimate a D-vine copula. The ordering $A_{j_d} - A_{j_{d-1}} - \dots - A_{j_1} - I$ is determined as outlined in Section \ref{DvineSetup}. Afterwards   $S$ conditional samples from the fitted conditional vine copula $C_{j_1,\dots,j_d|I}(\cdot|\alpha^I)$ on the copula scale can be simulated.
    \item In order to re-transform the copula scale residuals to forecasted logarithmic returns of the assets we use the marginal inverse distributions together with the forecasted mean and volatility estimates of the one step ahead time unit.
    \item Using the weights $w_j$ we can then aggregate the individual logarithmic return samples to portfolio level logarithmic return samples. For instance the market capitalization weights can be used.
    \item From the resulting vector of size $S$ we can derive the conditional risk measure estimates for the one step ahead time unit using the discussed Monte Carlo approach.
\end{itemize}

%

\FloatBarrier
\subsection{Rolling window}
\begin{figure}[!htbp]
    \centering
    \includegraphics[width = 0.65\textwidth]{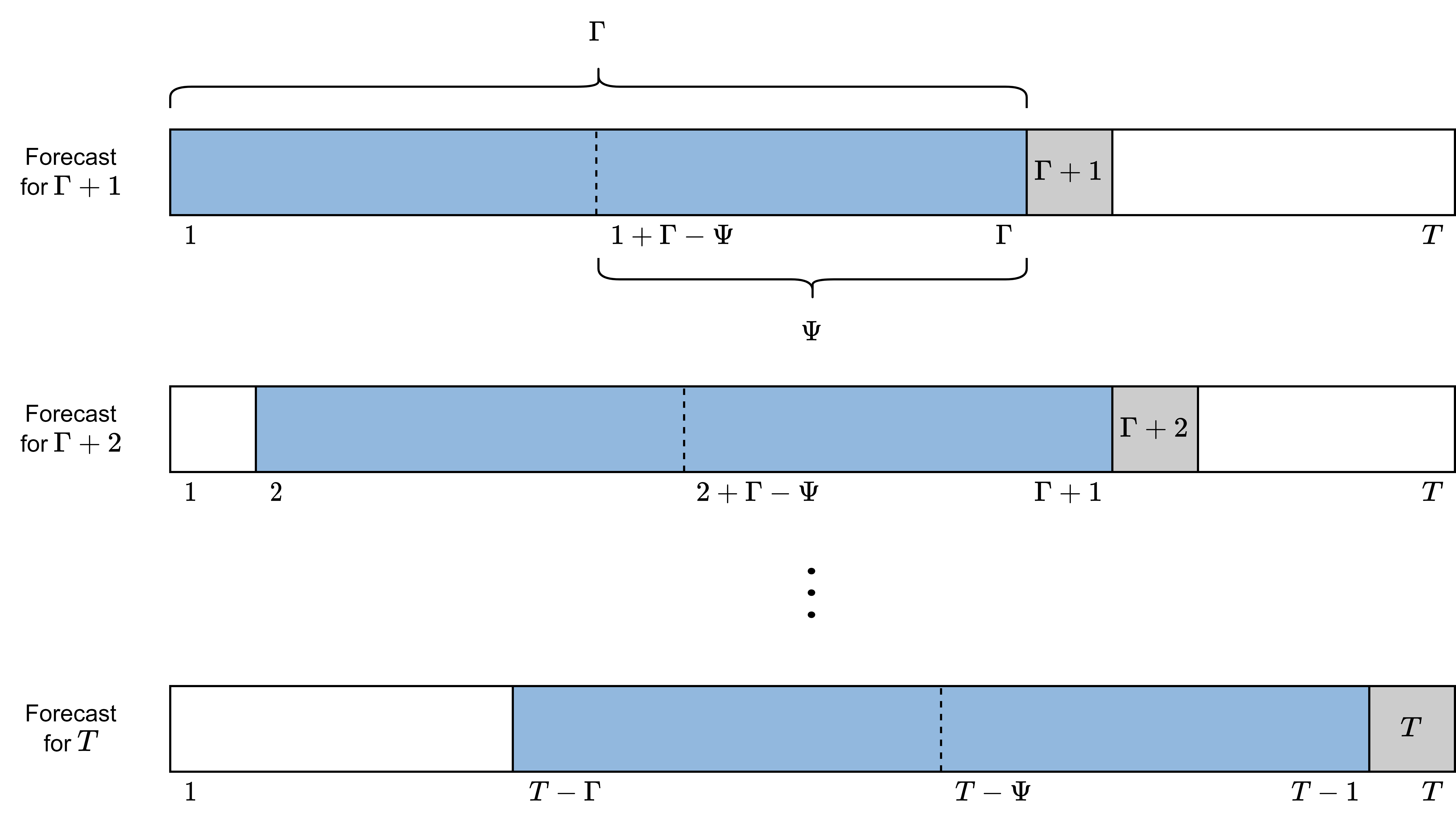}
    \caption{Illustration of the parameters $\Gamma$ and $\Psi$ alongside the one step ahead unconditional risk measure estimation.}
    \label{fig:onestep_rm_estimate_params}
\end{figure}

The one step forecasting approach as discussed in the previous section can then be applied repeatedly to estimate risk measures for the whole interval spanned by $\Gamma+1$ and $T$.  We implicitly assume here that the dependence structure assumed by the vine copula does not change for the upcoming one step ahead time unit. This repeated one step approach for the interval and the parameters $\Gamma$ and $\Psi$ is illustrated in Figure \ref{fig:onestep_rm_estimate_params}. 
Already for moderately large intervals of interest, like 100, a huge number of models has to be fitted which is a computational burden. Thus a rolling window approach is proposed with a refit of the models only after a certain time interval of usage. This of course might come to the price of potentially larger model misfits e.g. the dependence structure is assumed to stay constant over the longer intervals. 

For the rolling window approach we define the length of the usage window of the marginal models as  $\bm{\gamma} \leq (T-\Gamma) \in \mathbb{N}$ and the length of the usage window of the vine copula  model as  $\bm{\kappa} \leq \gamma \in \mathbb{N}$.
%

\begin{figure}
    \centering
    \includegraphics[width = 0.6\textwidth]{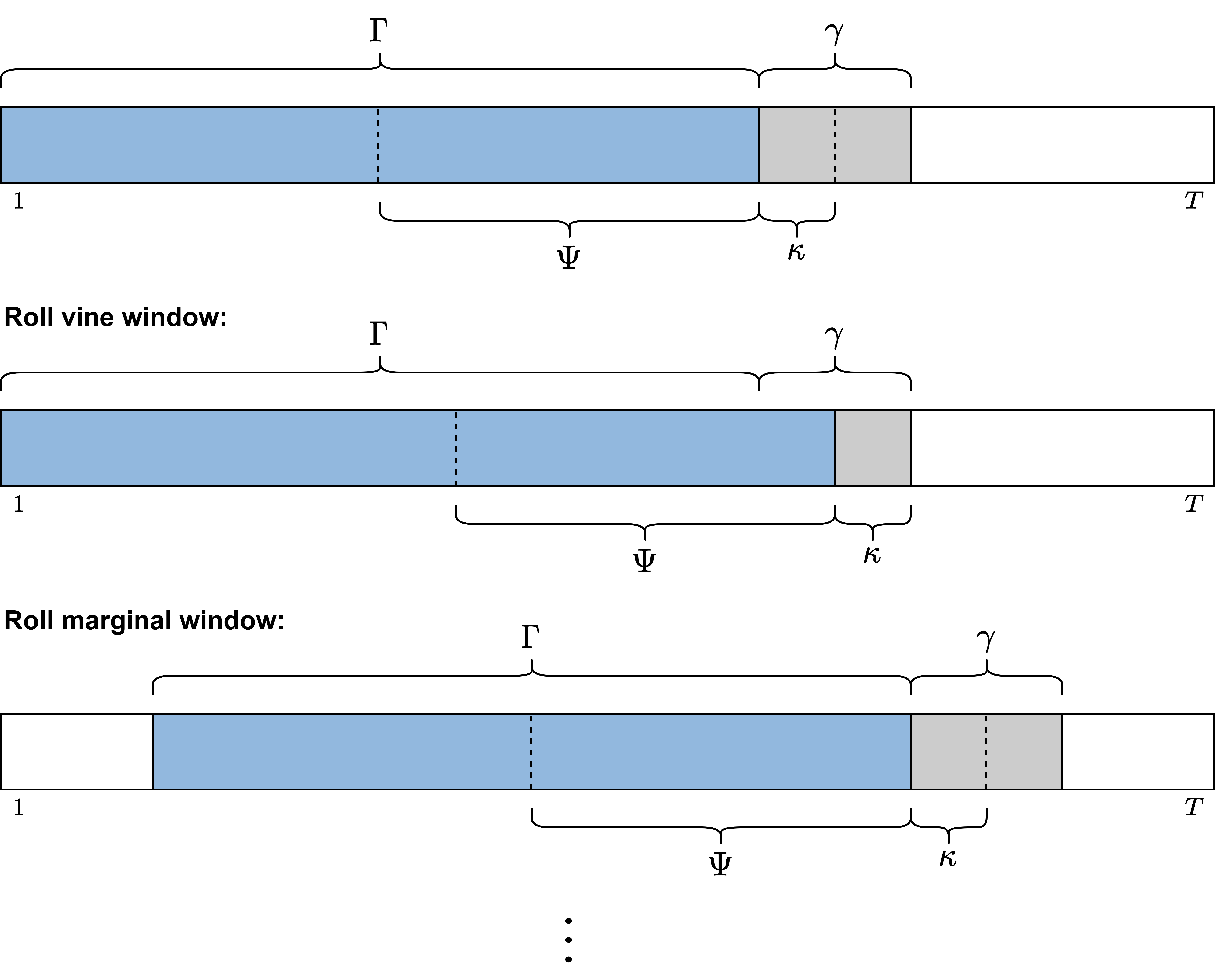}
    \caption{Illustration of the fitting windows ($\Gamma$ and $\Psi$) and the usage windows ($\gamma$ and $\kappa$) in the forecasting process.}
    \label{fig:rm_estimate_rollparams}
\end{figure}

It is reasonable to pick these parameters such that $(T-\Gamma) \bmod \gamma \equiv 0$ and $\gamma \bmod \kappa \equiv 0$ will have to hold. An illustration of the parameters and the rolling of the respective windows of the rolling window approach is given in Figure \ref{fig:rm_estimate_rollparams}.

Thus the resulting one step ahead conditional risk measure forecasting allows to analyze its robustness with respect to states of the overall or some specific market represented by $I$. These states are controlled via the $\alpha^I$ level. This might be applied to {stress testing} situations were we for example would like to assess the effect of a really bad performing market index $I$ or asset on the risk estimates. This corresponds in this setting to estimating conditional risk measures with a small $\alpha^I$ level. It might also be a more conservative risk measure for small $\alpha^I$ levels in the case of a generally positive dependence of the portfolio with the market index $I$. We will call this conditional strategy the {quantile based} one. In Section 7.2 of \cite{SommerThesis} we have also explored other strategies but this was the most promising one. \blue{At this point it should be further noted that the presented methodology is not restricted to a certain dimension of the portfolio and as vine copulas have a proven track record of capturing high dimensional dependencies this approach can be applied to high dimensional data. \cite{SommerThesis} elaborates on page 72 that testing was performed up to a maximum number of dimensions of 30 but just with the focus on accessing the scaling of the implementation towards higher dimensions.}

\FloatBarrier
\subsection{Software/Performance}
\sloppy
For the performance of the conditional risk measure estimation we  developed the \textit{portvine} package for the statistical computing language R, \citep{citeR} which is now publically available on CRAN. The thoroughly tested and documented package was build on the shoulders of giants, including the packages \textit{rvinecopulib} of \cite{Rvinecopulib} and \textit{rugarch} of \cite{rugarch}. These facilitate the effective estimation of vine copula and ARMA-GARCH models, respectively. 
It furthermore features extensive documentation with the main API documentation at \href{https://emanuelsommer.github.io/portvine/}{https://emanuelsommer.github.io/portvine/} and all package related references are given there.

The algorithms of the proposed risk measure estimation approaches are formulated to be highly parallelizable. In particular even nested parallel processing i.e. two levels of parallel processes are supported. This should definitely be exploited on modern tech stacks that often allow the use of many cores at once. Moreover the conditional sampling algorithms were implemented in the lower level programming language C++ which reduced the sampling runtime even further by a factor of roughly 30. This is important as especially for large sampling sizes (the parameter $S$) the runtime is mainly driven by the sampling.

In Table \ref{tab:cond_perf_scenarios} some runtime measurements that illustrates the significance of using the power of parallelization are shown. The measurements were performed on a Linux Cluster of the Leibniz Supercomputing Centre\footnote{https://www.lrz.de/services/compute/linux-cluster/}. The used R version is 3.6.0. For details on the extensive possibilities to set up the parallel strategy on your tech stack have a look at the package documentation and a more detailed performance assessment is given in \cite{SommerThesis}.

\begin{table}[hbt]
    \centering
    \begin{tabular}{cccc}
        $d$ & $S$ & Cores used & Runtime (minutes)  \\
        \hline
        10 & 10000 & 20 & 24.11 \\
        10 & 10000 & 250 & 6.22 \\
        30 & 10000 & 20 & 119.25 \\
        10 & 100000 & 20 & 231.20 \\
        10 & 100000 & 250 & 34.67 (only 1 quantile level) \\
        \hline
    \end{tabular}
    \caption{Overview of the runtime of some of the scenarios from the performance assessment of the conditional risk measure estimation approach. In all displayed cases all parametric copula families were allowed, the observations per variable were fixed to $T=1000$ and 5 marginal windows of size 50 as well as 10 vine windows of size 25 were used. The vine training set size was $\Psi = 200$ and in every case both the VaR and ES were estimated at two $\alpha$ levels each. Besides the case in the fifth scenario, which would require RAM of $>$100Gb for two quantile levels, always two quantile levels $\alpha^I$ were estimated.}
    \label{tab:cond_perf_scenarios}
\end{table}

\section{Case Study}
\begin{figure}[hbt!]
\centering
\includegraphics[width=\textwidth]{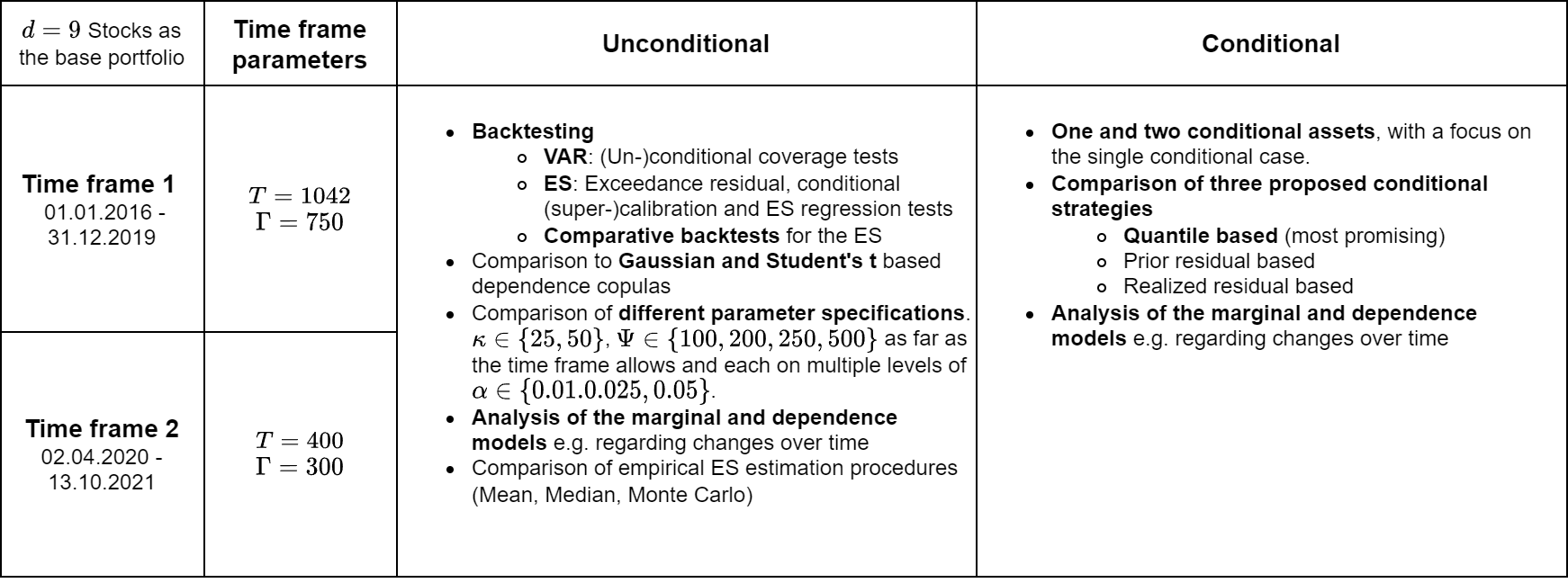}
\caption{An overview of the main topics covered in the case studies featured in \cite{SommerThesis}.}
\label{fig:casestudyintro}
\end{figure}

\noindent In the following we provide \blue{an exemplary} case study performing the conditional risk measure estimation approach using the newly designed \textit{portvine} package \footnote{The source code for the upcoming analyses is also publicly available on GitHub at \href{https://github.com/EmanuelSommer/PortvineThesis}{https://github.com/EmanuelSommer/PortvineThesis}.}. \blue{This is only an excerpt of much more exhaustive case studies presented in \cite{SommerThesis}. An overview of the main topics covered in those can be found in Figure \ref{fig:casestudyintro}. The comparisons in the unconditional case were performed using the comparative backtests introduced in \cite{noldeESbacktest} and are discussed for this application in Section 8.2 of \cite{SommerThesis}.}

\blue{In addition, we have added a small proof of concept simulation study in the \ref{app:simstudy}. There we examine the performance of the unconditional approach under the assumption that the ARMA-GARCH models can adequately model the marginals and that the dependency structure can be estimated via a vine copula. The obtained results do not indicate any structural failure of the approach in the unconditional case. Due to the same set of assumptions and the rigorous derivation of the conditional sampling from the D-vine via the Rosenblatt transform the results should translate very well to the conditional case.
}

\subsection{Data and Setup}
We consider a portfolio of nine Spanish stocks $d = 9$ which are the  top constituents of the MSCI Spain stock market index over \blue{the time frame} of 02.04.2020 - 13.10.2021 ($T = 400$ daily log returns and a training window of $\Gamma = 300$). The nine assets along their corresponding weights $w_j$ are given in Table \ref{tab:marketcap}. The weight corresponds to the respective market capitalization (EUR billions) of the asset on the 29th October 2021. \blue{As indicated in Figure \ref{fig:casestudyintro} a case study with the same setup but a longer time frame of $T = 1042$ and $\Gamma = 750$ days is given in detail in Section 10 of \cite{SommerThesis}. The results however are matching to the ones presented here and thus the higher volatility time frame during the pandemic was selected to be displayed here.}

The additional two conditional assets that will be considered are the Standard's and Poor 500 (S\&P 500) index and the Eurostoxx 50 index. As our portfolio consists of Spanish stocks we expect the influence of the Eurostoxx 50 index to be greater than the one of the S\&P 500. Some assets are also part of the Eurostoxx index, namely the Banco Santander, the Amadeus It Group, Iberdrola and Inditex. However, they cumulatively only account for roughly 5\% of the index.


In this case study, we choose a  forecasting and usage window size for the marginal time series models of $\gamma, \kappa = 50$ which corresponds to roughly 2 financial months and results in 2 marginal refits. The default marginal model is an ARMA(1,1)-GARCH(1,1) model with the residual white noise distribution being a skewed Student's t distribution. The assessment of the quality of marginal model fit did not indicate the need to deviate from the default marginal model in any of the cases. \blue{
An important part of this assessment is the check of the approximation of the i.i.d. innovations by their estimate the standardized residuals. As shown in \cite{francqzakonian} under consistent estimation, as is the case for univariate ARMA-GARCH models, the standardized residuals converge towards the innovations. Moreover our approach only requires in
the first step univariate time series models and then models in a second step the dependency on the derived copula scale, therefore we do not need the joint i.i.d. property as investigated in \cite{serialdepresiduals}. This two step approach is an extension of the approach commonly followed in copula applications first suggested by \cite{Joe_Xu_1996}. They also proofed the standard asymptotic properties for the case considered there. For the assessment whether this approximation is reasonable we have employed among other tests univariate Ljung-Box tests.
} The full residual analysis is featured in Chapter 10 in  \cite{SommerThesis}.

For the D-vine fitting all parametric bivariate copula families implemented in the \textit{rvinecopulib} package of \cite{Rvinecopulib} were allowed. For all models the VaR and the ES are estimated for the $\alpha$ levels 0.01, 0.025, 0.05.  Moreover all estimated risk measures are based on a realistic simulation sample size of $S = 100000$ samples.
The parameters concerning the D-vine copula models will be fixed to $\Psi = 200$ (fitting window) and $\kappa = 50$ (usage window). \blue{These window sizes were suggested by comparative backtests on page 101 of \cite{SommerThesis}.} The focus during this conditional analysis will lie on the ES estimated at the level $\alpha = 0.025$.

\subsection{Comparison S\&P 500 and EuroStoxx 50 as conditional assets}

\begin{figure}[hbt!]
     \centering
     \begin{subfigure}[b]{0.7\textwidth}
         \centering
        \includegraphics[width = \textwidth]{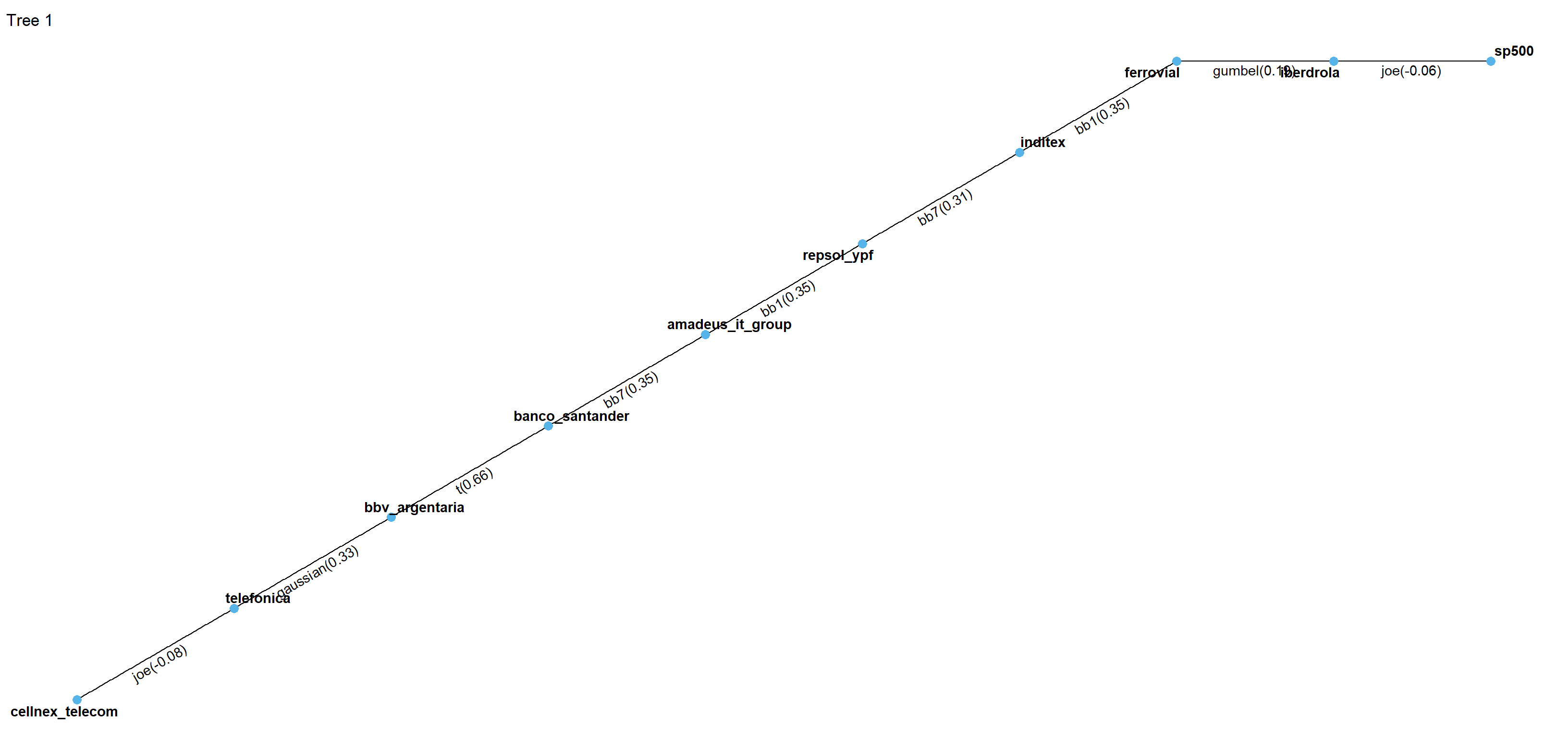}
        \caption{First vine usage window when conditional asset is the {S\&P 500}.}
        \label{fig:dvines_cond1_20sp500_1}
     \end{subfigure}

     \begin{subfigure}[b]{0.7\textwidth}
         \centering
           \includegraphics[width = \textwidth]{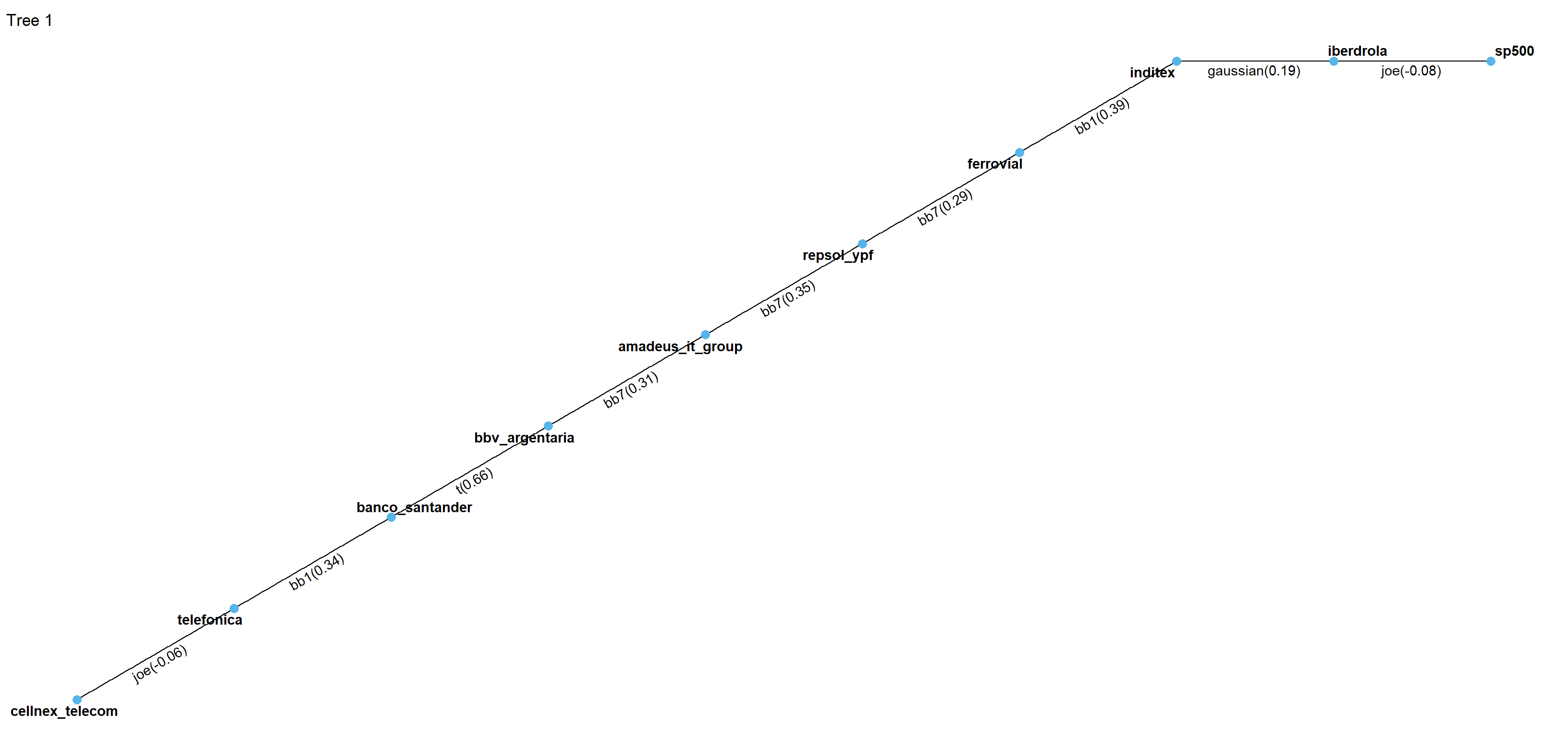}
    \caption{Second vine usage window when conditional asset is the S\&P 500.}
    \label{fig:dvines_cond1_20sp500_2}
     \end{subfigure}

     \begin{subfigure}[b]{0.7\textwidth}
         \centering
    \includegraphics[width = \textwidth]{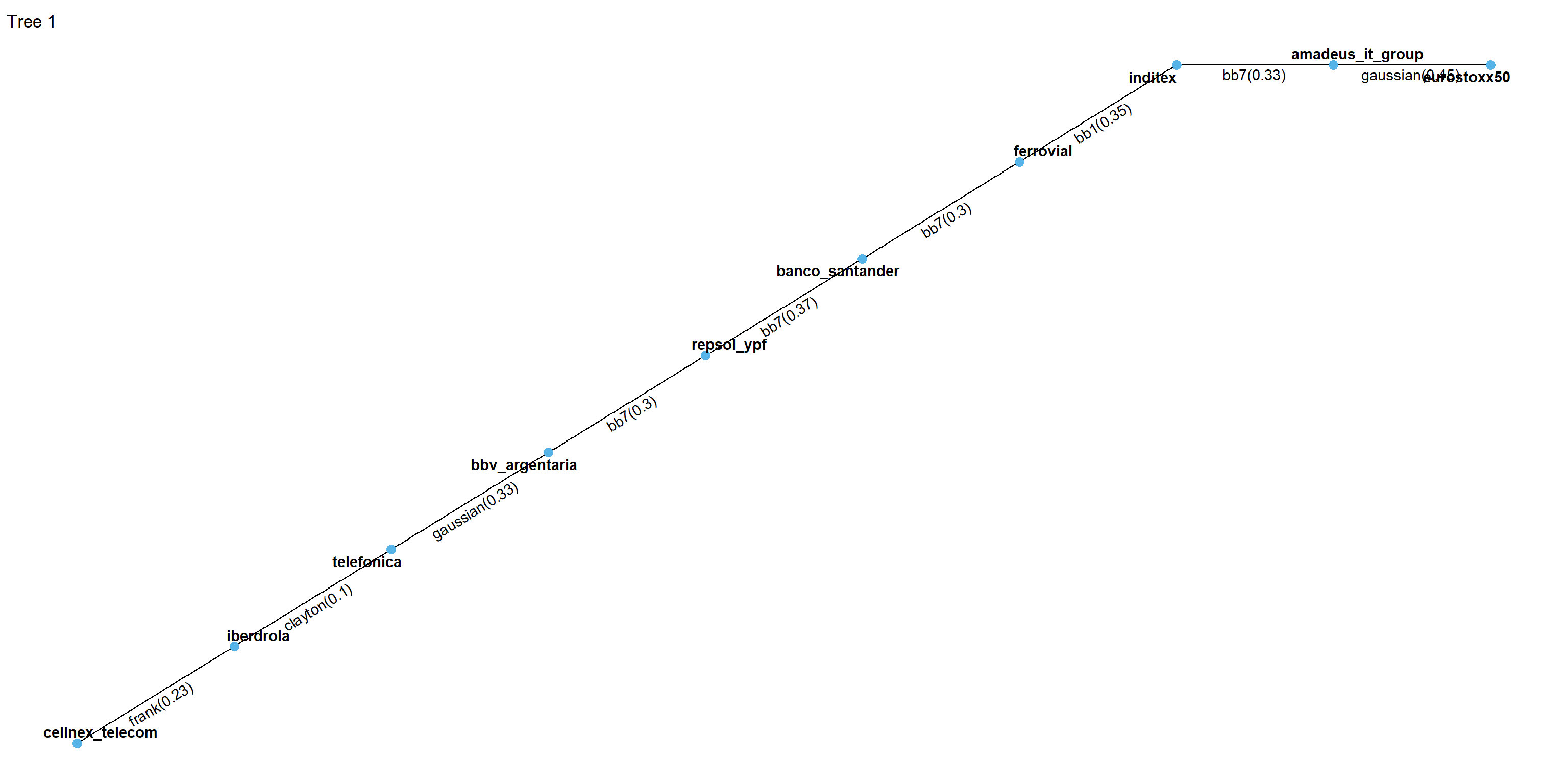}
    \caption{First vine usage window when conditional asset is the {Eurostoxx 50}.}
    \label{fig:dvines_cond1_20euro_1}
     \end{subfigure}

     \begin{subfigure}[b]{0.7\textwidth}
         \centering
    \includegraphics[width = \textwidth]{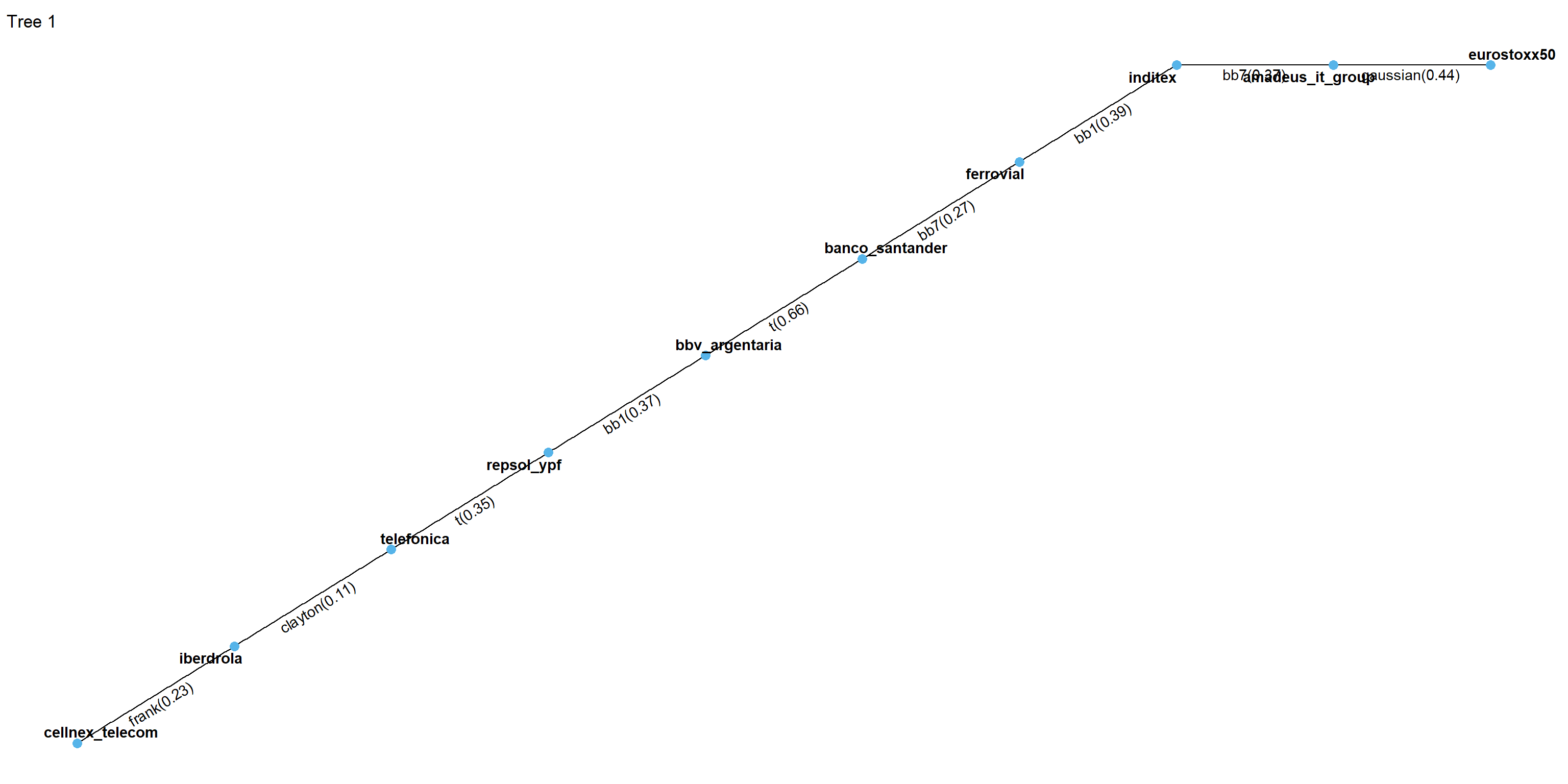}
    \caption{Second vine usage window when conditional asset is the {Eurostoxx 50}.}
    \label{fig:dvines_cond1_20euro_2}
     \end{subfigure}
        \caption{First  D-vine tree  with chosen pair copula family and  Kendall`s $\tau$ in parenthesis. }
        \label{fig:three graphs}
\end{figure}

Now we fit a D-vine for each index as with the index as last node.  Comparing the vine trees, we find that the  S\&P 500 shows weak dependence between the  the conditional S\&P 500 asset and the rightmost portfolio asset in all cases as shown in Figure  \ref{fig:dvines_cond1_20sp500_1} and Figure \ref{fig:dvines_cond1_20sp500_2}.  Although the American market is one of the most important export markets outside of
Europe for Spain, the index seems to have very little influence on our portfolio. Consequently also the ordering of the other stocks changes very strongly over the two usage windows. In addition we observe that many independence copulas (Table \ref{tab:bicopsused20cond_sp500}) are chosen and no influence of the conditioning variable in higher tree levels is found(Table \ref{tab:bicopsused20cond_sp500right}).

Following the same setup but using the Eurostoxx 50 instead of the S\&P 500 as the outer leaf, we find quite strong positive bivariate dependence
between the conditional asset the Eurostoxx 50 index and the rightmost portfolio asset which is always the  Amadeus It Group as shown in Figure \ref{fig:dvines_cond1_20euro_1} and Figure \ref{fig:dvines_cond1_20euro_2}.Also the ordering of the assets is quite stable and less independence copulas are chosen  (Table \ref{tab:bicopsused20cond_euro}). This indicates that the conditional risk measures based on this dependence structure will be quite strongly influenced by the conditional asset.

\begin{figure}[hbt!]
     \centering
     \begin{subfigure}[b]{0.8\textwidth}
         \centering
    \includegraphics[width = 0.9\textwidth]{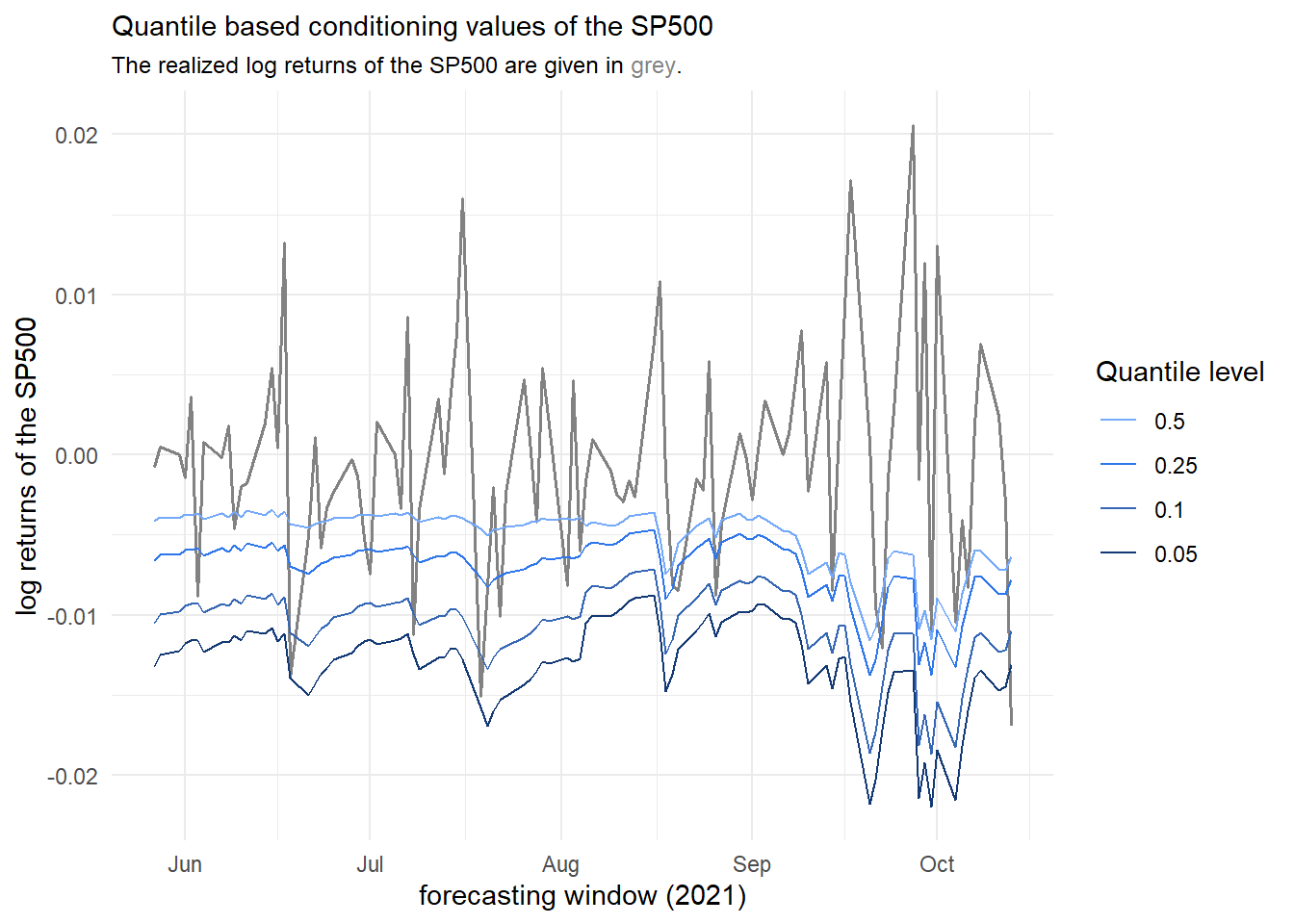}
    \caption{The conditioning asset is {S\&P 500}.}
    \label{fig:cond1_sp500_20_quantiles}
     \end{subfigure}

     \begin{subfigure}[b]{0.8\textwidth}
         \centering
            \includegraphics[width = 0.9\textwidth]{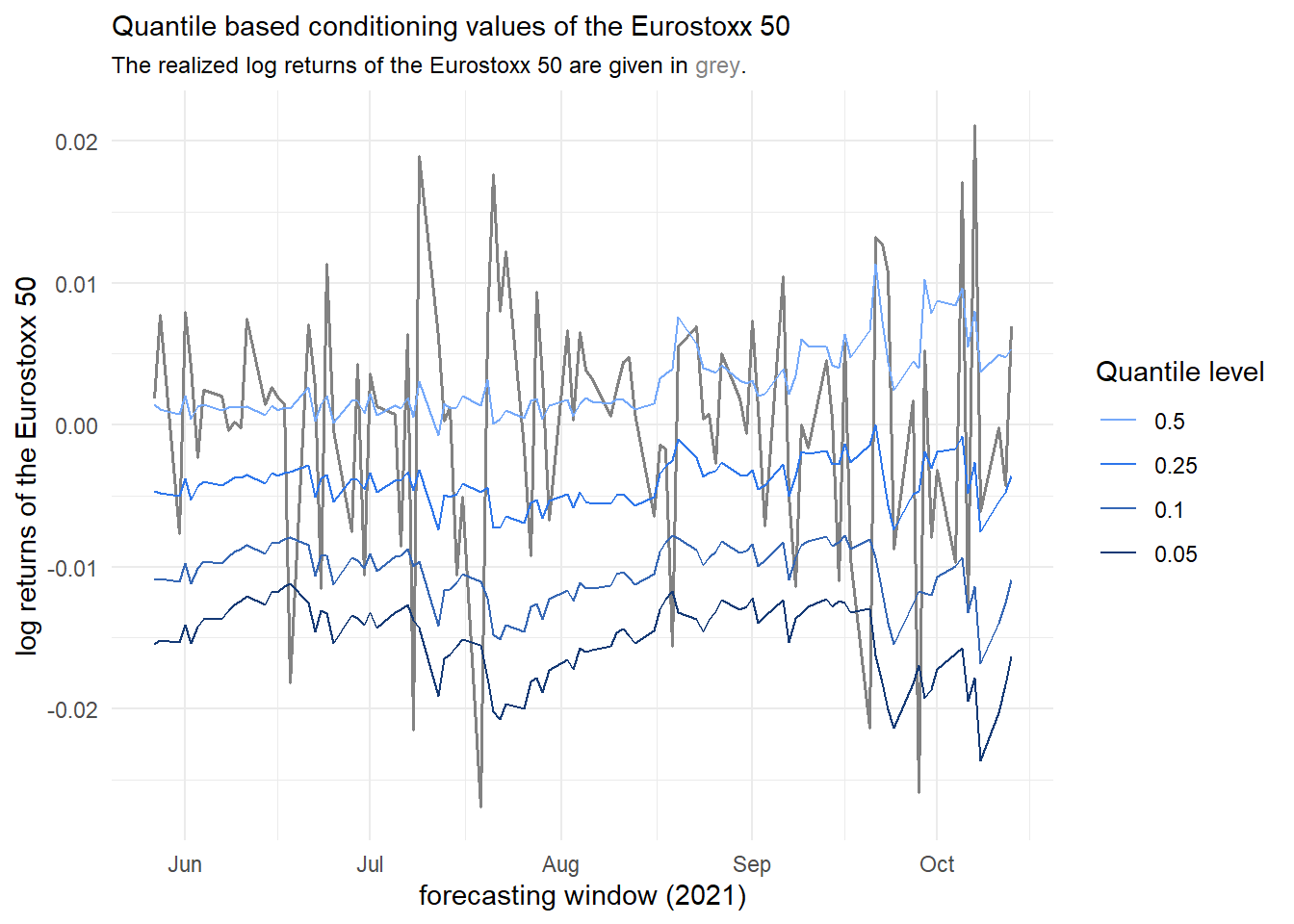}
    \caption{The conditioning asset is Eurostoxx 50.}
    \label{fig:cond1_euro_20_quantiles}
     \end{subfigure}
        \caption{Behavior of the {quantile based conditioning values} of different conditioning assets.}
        \label{fig:cond1}
\end{figure}

Next, we are focusing on the conditioning values for the indices. Using different quantile levels $\alpha^I$ on the copula scale, we can express a bad performance of the conditional asset with values close to 0 and a good performance over the time frame of interest with values closer to 1.  By re-transforming this conditional copula quantile level by the inverse marginal distribution function, we obtain conditional values on the original log return scale.  We visualize these conditioning values on the log return scale alongside the realized time series of the conditional asset. This allows us to see whether the quantile based conditioning values actually do what we expect. Figures \ref{fig:cond1_sp500_20_quantiles} and \ref{fig:cond1_euro_20_quantiles} show the quantile based conditioning values for both considered conditional assets. In both cases we can observe that the quantile based approach produces forecasted time series that are expressing various states of the conditioning asset.

Now we visualize the conditional risk measure estimates and compare them to the corresponding unconditional estimates. We focus on the ES forcasted for the level  $\alpha = 0.025$ for the portfolio.

\begin{figure}[hbt!]
     \centering
     \begin{subfigure}[b]{0.8\textwidth}
         \centering
    \includegraphics[width=\textwidth]{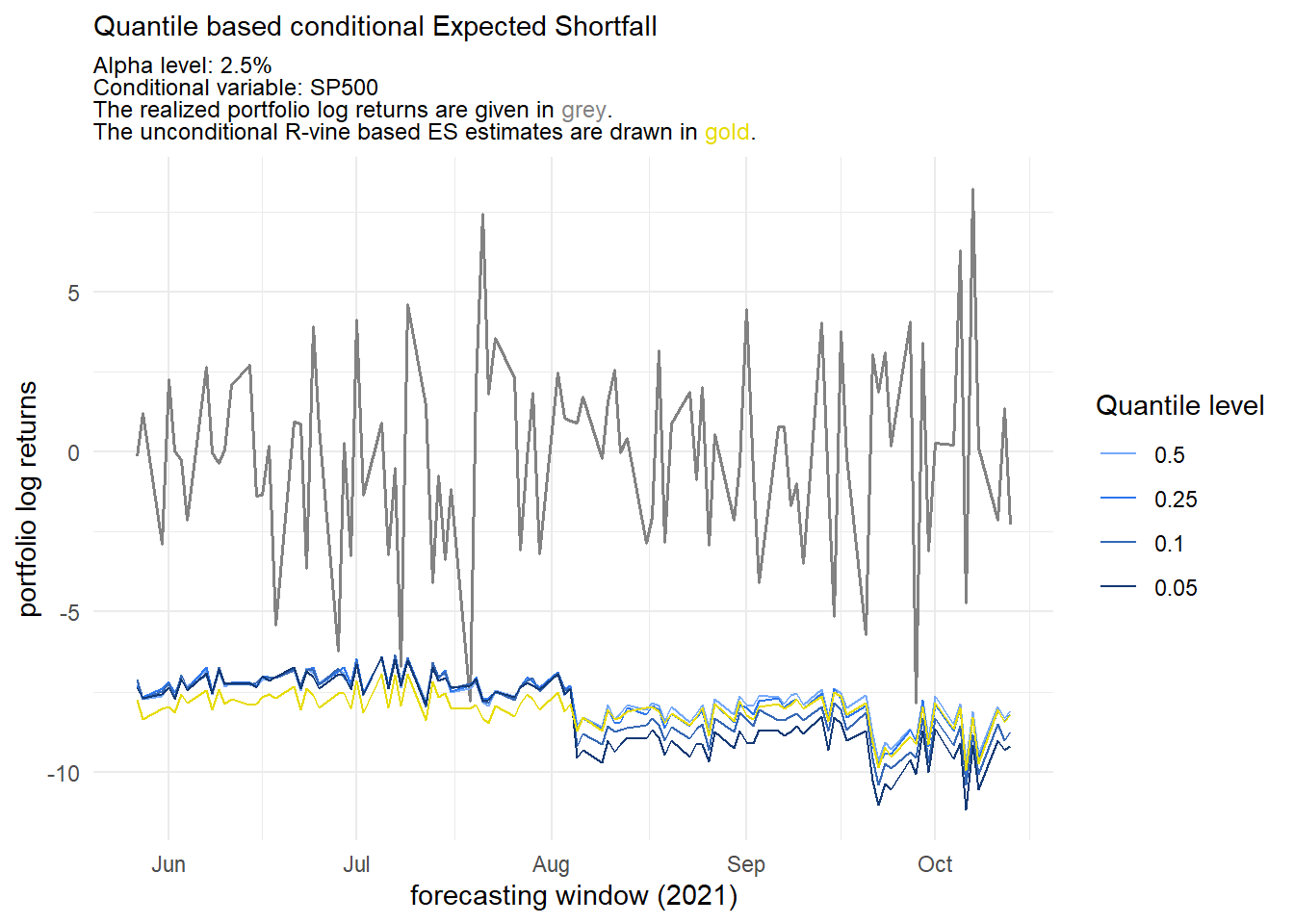}
    \caption{The conditioning asset is the SP500.}
    \label{fig:cond1_rm_sp500_20_quantiles3}
     \end{subfigure}

     \begin{subfigure}[b]{0.8\textwidth}
         \centering
    \includegraphics[width=\textwidth]{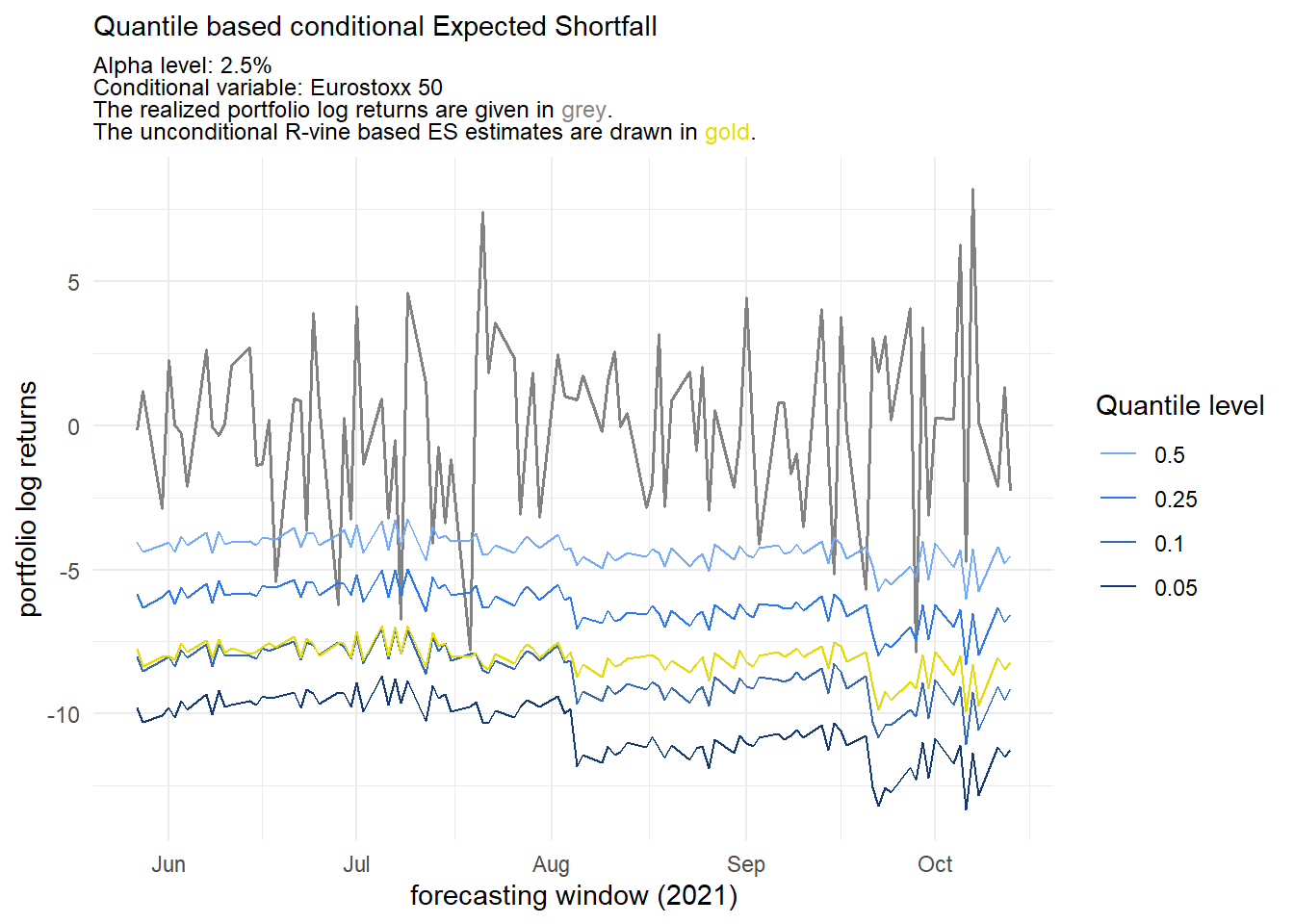}
    \caption{The conditioning asset is the Eurostoxx 50.}
    \label{fig:cond1_rm_euro_20_quantiles3}
     \end{subfigure}
        \caption{Comparison of the {quantile based conditional ES forecasts} at $\alpha = 0.025$ (blue) with the {unconditional R-vine based ES estimates} (gold) with different condition assets.}
        \label{fig:cond1_rm}
\end{figure}

Focusing on the S\&P 500,  we find that the resulting one day ahead forcasted conditional ES series lie quite close to each other (Figure \ref{fig:cond1_rm_sp500_20_quantiles3}) which was already anticipated by the weak dependence in the corresponding D-vines. As the portfolio level ES seems to be quite robust against strong market downtrends of the American market one could argue that the portfolio might be a good hedge against the risk of a downtrend in the American market. 

For the Eurostoxx 50 index the different series corresponding to the quantile levels are clearly distinguishable  (Figure \ref{fig:cond1_rm_euro_20_quantiles3}) and thus states of the European economy have a serious effect on the estimated ES. The most interesting observation we can make here is that the unconditional ES series is close to the one corresponding to the quantile level $\alpha^I = 0.1$. Thus we can argue that by using only the unconditional approach to estimate the ES at the
confidence level $\alpha = 0.025$ we have already accounted for a quite negative general European market. However if we would like to account for an even worse European market for example at the quantile level $0.05$ we would have to further adjust our risk estimates to be more conservative as indicated by the series corresponding to the quantile level $\alpha^I = 0.05$. \blue{This assessment of of sensitivity of the risk measures against an external stressor is critical for decision makers to understand how robust the basic unconditional risk measure estimation is. This means certain critical dependencies can be tested and potentially underestimated risk exposures might be uncovered.} These conditional risk measure estimates can then also be used to further analyze the effects on capital requirements or other financial key metrics.

If one is interested in using two conditional indices, this has also been implemented in the \textit{portvine} package and showcased in \cite{SommerThesis}.

\FloatBarrier
\section{Conclusion and Outlook}

In this paper, we focus on extending the estimation of the unconditional risk measure by introducing a novel algorithm for forecasting the conditional risk measure with a focus on the ES.  By doing so, we are able to effectively measure the risk of potentially large asset portfolios in a conditional setting using a stress factor.  This work can benefit investors as well as regulators and financial institutions by providing more accurate access to the estimation of various risk measures of financial portfolios. \blue{Conditional risk measures allow sensitivity analyses of the estimated risk measures to stress by an other asset or market indicator.}
We use the flexible class of vine copulas, more specifically the D-vine copulas, to determine the conditional distribution of the components given a stress factor and to model the complex dependence structures. This allows the \blue{potentially} high dimensional dependence structure of the financial portfolio to be adequately incorporated.

We have presented here an exemplary case study on Spanish equities to show how this novel algorithm can be applied.  Using the European and American market index as a stress factor, we find that a portfolio of the largest Spanish stocks could potentially be used as a hedge in the American market, while this is not possible in the case of the European market. Here, the sensitivity of the risk measure to the market is much higher for Spanish assets. \blue{While we have presented here a small proof-of-concept simulation study, a full-scale simulation study is currently underway, highlighting that based on the work presented, many interesting research questions have arisen, such as appropriate backtests for the conditional risk measure series.}

 We have also contributed by developing a novel algorithm to sample from the stress factors of a D-vine copula. The developed computationally efficient R package \textit{portvine} is able to model both unconditional and conditional risk measures. It is open source and available on CRAN.

\section{Acknowledgments}
The provided computational resources by the  Linux Cluster of the Leibniz Supercomputing Centre (\url{www.lrz.de}) are gratefully acknowledged. Claudia Czado is supported by the German Research Foundation (DFG). \blue{Furthermore we are thankful to two anonymous referees helping us to further improve the manuscript.}

\clearpage
\appendix
\section{Additional Information Case Study}

\begin{table}[hbt]
    \centering
    \begin{tabular}{cccc}
        Asset & Iberdrola & Banco Santander & BBV Argentaria \\
        \hline
        Weight  & 60.48 & 56.82 & 40.42  \\
        \hline \hline
        Asset  & Inditex & Cellnex Telecom & Amadeus It Group  \\
        \hline
        Weight  & 34.08 & 27.09 & 26.06 \\
        \hline \hline
        Asset &  Telefonica & Repsol Ypf & Ferrovial  \\
        \hline
        Weight  & 19.37 & 16.04 & 12.99  
    \end{tabular}
    \caption{The assets of the Spanish stock portfolio of interest with their respective market capitalization (EUR Billions) on the 29.10.2021.}
    \label{tab:marketcap}
\end{table}

\begin{table}[hbt!]
\tiny
\begin{tabular}{cccccc}
\begin{tabular}[c]{@{}c@{}}Copula\\  family\end{tabular} & \begin{tabular}[c]{@{}c@{}}Count\\ 1. usage window\end{tabular} & \begin{tabular}[c]{@{}c@{}}Count\\ 2. usage window\end{tabular} & \begin{tabular}[c]{@{}c@{}}Copula \\ family\end{tabular} & \begin{tabular}[c]{@{}c@{}}Count\\ 1. usage window\end{tabular} & \begin{tabular}[c]{@{}c@{}}Count\\ 2. usage window\end{tabular} \\ \hline
Gaussian                                                 & 2                                                               & 5                                                               & Independence                                             & 12                                                              & 10                                                              \\
Student's t                                              & 4                                                               & 5                                                               & Clayton                                                  & 5                                                               & 2                                                               \\
Frank                                                    & 6                                                               & 7                                                               & Gumbel                                                   & 2                                                               & 4                                                               \\
Joe                                                      & 7                                                               & 6                                                               & BB1                                                      & 3                                                               & 2                                                               \\
BB7                                                      & 3 & 3 & BB8                                                      & 1                                                               & 1
\end{tabular}
    \caption{The used copula families alongside their frequencies for the fitted D-vine of the first and second usage window. The conditional asset is the SP500.}
      \label{tab:bicopsused20cond_sp500}
\end{table}

\begin{table}[hbt]
\centering
    \begin{tabular}{cccc}
        Tree level & Copula & Copula family & Empirical Kendall's tau  \\
        \hline
        1 & $C_{j_1,I}$ & Joe & -0.0566 \\
        2 & $C_{j_2,I; j_1}$ & Independence & 0\\
        3 & $C_{j_3,I; j_1, j_2}$ & Joe & -0.0675 \\
        4 & $C_{j_4,I; j_1:j_3}$ & Joe & 0.0972 \\
        5 & $C_{j_5,I; j_1:j_4}$ & Independence & 0\\
        6 & $C_{j_6,I; j_1:j_5}$ & Independence & 0\\
        7 & $C_{j_7,I; j_1:j_6}$ & Independence & 0\\
        8 & $C_{j_8,I; j_1:j_7}$ & Independence & 0\\
        9 & $C_{j_9,I; j_1:j_8}$ & Independence & 0
        \end{tabular}
    \caption{The used copula families alongside their empirical Kendall's tau associated with the conditional asset for the fitted D-vine of the first usage window. The conditional asset is the SP500. ($j_1$ = Iberdrola, $j_2$ = Ferrovial, $j_3$ = Inditex, $j_4$ = Repsol Ypf, $j_5$ = Amadeus It Group, $j_6$ = Banco Santander, $j_7$ = BBV Argentaria, $j_8$ = Telefonica, $j_9$ = Cellnex Telecom)}
    \label{tab:bicopsused20cond_sp500right}
\end{table}

\begin{table}[hbt!]
\tiny
\begin{tabular}{cccccc}
\begin{tabular}[c]{@{}c@{}}Copula\\  family\end{tabular} & \begin{tabular}[c]{@{}c@{}}Count\\ 1. usage window\end{tabular} & \begin{tabular}[c]{@{}c@{}}Count\\ 2. usage window\end{tabular} & \begin{tabular}[c]{@{}c@{}}Copula \\ family\end{tabular} & \begin{tabular}[c]{@{}c@{}}Count\\ 1. usage window\end{tabular} & \begin{tabular}[c]{@{}c@{}}Count\\ 2. usage window\end{tabular} \\ \hline
Gaussian                                                                                                                         & 5                                                                                                                                       & 5                                                                                                                                       & Independence                                                                                                                     & 6                                                                                                                                       & 7                                                                                                                                       \\
Student's t                                                                                                                      & 4                                                                                                                                       & 5                                                                                                                                       & Clayton                                                                                                                          & 2                                                                                                                                       & 5                                                                                                                                       \\
Frank                                                                                                                            & 11                                                                                                                                      & 6                                                                                                                                       & Gumbel                                                                                                                           & 6                                                                                                                                       & 6                                                                                                                                       \\
Joe                                                                                                                              & 5                                                                                                                                       & 4                                                                                                                                      & BB1                                                                                                                              &1                                                                                                                                     & 2                                                                                                                                       \\
BB7                                                                                                                              & 5                                                                                                                                       & 4                                                                                                                                       & BB8                                                                                                                              & -                                                                                                                                       & 1                                                                                                                                      
\end{tabular}
    \caption{The used copula families alongside their frequencies for the first and second fitted D-vine. The conditional asset is the Eurostoxx 50.}
    \label{tab:bicopsused20cond_euro}
\end{table}

\begin{table}[hbt]
\centering
    \begin{tabular}{cccc}
        Tree level & Copula & Copula family & Empirical Kendall's tau  \\
        \hline
        1 & $C_{j_1,I}$ & Gaussian & 0.4461 \\
        2 & $C_{j_2,I;j_1}$ & Frank & 0.3078\\
        3 & $C_{j_3,I; j_1,j_2}$ & Frank & 0.1747 \\
        4 & $C_{j_4,I; j_1:j_3}$ & Gumbel & 0.1528 \\
        5 & $C_{j_5,I; j_1:j_4}$ & Independence & 0\\
        6 & $C_{j_6,I; j_1:j_5}$ & Gaussian & -0.0902\\
        7 & $C_{j_7,I; j_1:j_6}$ & Independence & 0\\
        8 & $C_{j_8,I; j_1:j_7}$ & Frank & 0.3420\\
        9 & $C_{j_9,I; j_1:j_8}$ & Joe & 0.0404
        \end{tabular}
    \caption{The used copula families alongside their empirical Kendall's tau associated with the conditional asset for the fitted D-vine of the first usage window. The conditional asset is the Eurostoxx 50. ($j_1$ = Amadeus It Group, $j_2$ = Inditex, $j_3$ = Ferrovial, $j_4$ = Banco Santander, $j_5$ = Repsol Ypf, $j_6$ = BBV Argentaria, $j_7$ = Telefonica, $j_8$ = Iberdrola, $j_9$ = Cellnex Telecom)}
    \label{tab:bicopsused20cond_euroright}
\end{table}

\clearpage
\section{Proof of Concept Simulation Study (unconditional)}
\label{app:simstudy}
\blue{
In the following PoC Simulation Study we want to showcase that given the main assumptions of the approach we can estimate the portfolio risk measures as intended. We do this for the unconditional setting. As the conditional simulation from the D-vine is derived rigorously from the Rosenblatt transform in \cite{SommerThesis} and thus does not involve approximations the results from this simulation should translate very well to the conditional case.}

\begin{figure}[hbt!]
\centering
\includegraphics[width=\textwidth]{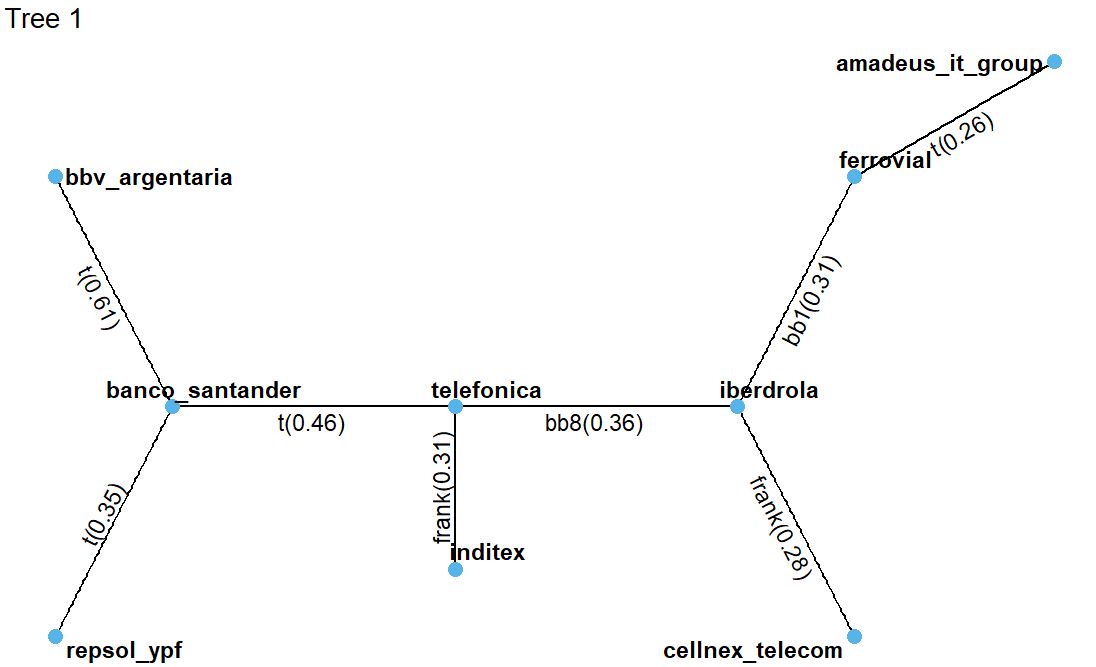}
\caption{The first tree of the fixed R-vine used for the simulation as the true series. It corresponds to the one fitted for the first vine window of the case study with the time frame 01.01.2016 to the 31.12.2019 described in section 10 of \cite{SommerThesis}.}
\label{fig:simstudyfirsttree}
\end{figure}

\blue{As a starting point we use data and fitted marginal and vine copula models from the first rolling window of the analysis of Spanish stocks for the time frame from the 01.01.2016 to the 31.12.2019 described in detail in Section 10 of \cite{SommerThesis}. The estimated parameters of the marginal models are reported in Table \ref{tab:simstudyarmagarch} and the first tree level of the fitted R-vine is displayed in Figure \ref{fig:simstudyfirsttree}. }

\begin{table}[hbt]
    \centering
    \begin{tabular}{ccccc}
        Asset & $\phi_0$ & $\phi_1$ & $\theta_1$ & $\alpha_0$ \\
        \hline
        Iberdrola & $1.273 \times 10^{-4}$ & $0.699$ & $-0.672$ & $2.00 \times 10^{-5}$ \\
        Banco Santander & $-1.138 \times 10^{-4}$ & $0.313$ & $-0.301$ & $7.102 \times 10^{-6}$ \\
        Inditex & $-1.389 \times 10^{-4}$ & $-0.145$ & $0.119$ & $2.462 \times 10^{-6}$ \\
        Cellnex Telekom & $4.865 \times 10^{-4}$ & $0.180$ & $-0.124$ & $2.574 \times 10^{-5}$ \\
        Respol YPF & $6.784 \times 10^{-4}$ & $-0.366$ & $0.411$ & $2.700 \times 10^{-6}$ \\
        Ferrovial & $1.603 \times 10^{-4}$ & $0.167$ & $-0.115$ & $4.840 \times 10^{-5}$ \\
        Amadeus It Group & $9.288 \times 10^{-4}$ & $-0.734$ & $0.775$ & $1.190 \times 10^{-5}$ \\
        Telefonica & $-3.716 \times 10^{-4}$ & $-0.268$ & $0.328$ & $8.874 \times 10^{-6}$ \\
        BBV Argentaria & $-2.503 \times 10^{-4}$ & $-0.014$ & $0.063$ & $9.724 \times 10^{-6}$ \\
        \hline
        Asset & $\alpha_1$  & $\beta_1$ & $\lambda$ & $\nu$ \\
        \hline
        Iberdrola & $0.140$ & $0.699$ & $1.000$ & $4.973$ \\
        Banco Santander & $0.062$ & $0.915$ & $0.974$ & $5.819$ \\
        Inditex & $0.030$ & $0.955$ & $1.077$ & $5.164$ \\
        Cellnex Telekom & $0.078$ & $0.790$ & $1.048$ & $5.935$ \\
        Respol YPF & $0.064$ & $0.925$ & $0.991$ & $5.209$ \\
        Ferrovial & $0.163$ & $0.545$ & $1.072$ & $5.349$ \\
        Amadeus It Group & $0.086$ & $0.838$ & $0.941$ & $4.384$ \\
        Telefonica & $0.119$ & $0.846$ & $1.024$ & $4.540$ \\
        BBV Argentaria & $0.074$ & $0.897$ & $1.014$ & $5.619$ \\
    \end{tabular}
    \caption{The fixed Coefficients of the ARMA(1,1)-GARCH(1,1) model used for the simulation as the true series. They correspond to the marginal models fitted for the first marginal window of the case study with the time frame 01.01.2016 to the 31.12.2019 described in Section 10 of \cite{SommerThesis}. The mean model is parameterized by the intercept $\phi_0$, the AR ($\phi_1$) and the MA ($\theta_1$) coefficients. The variance model respectively is parameterized by the variance intercept $\alpha_0$, the ARCH ($\alpha_1$) and the GARCH ($\beta_1$) coefficients. The residual distribution was in every case a skewed Student`s t distribution with the estimated skew ($\lambda$) and shape ($\nu$) parameters.}
    \label{tab:simstudyarmagarch}
\end{table}

\begin{figure}[hbt!]
\centering
\includegraphics[width=\textwidth]{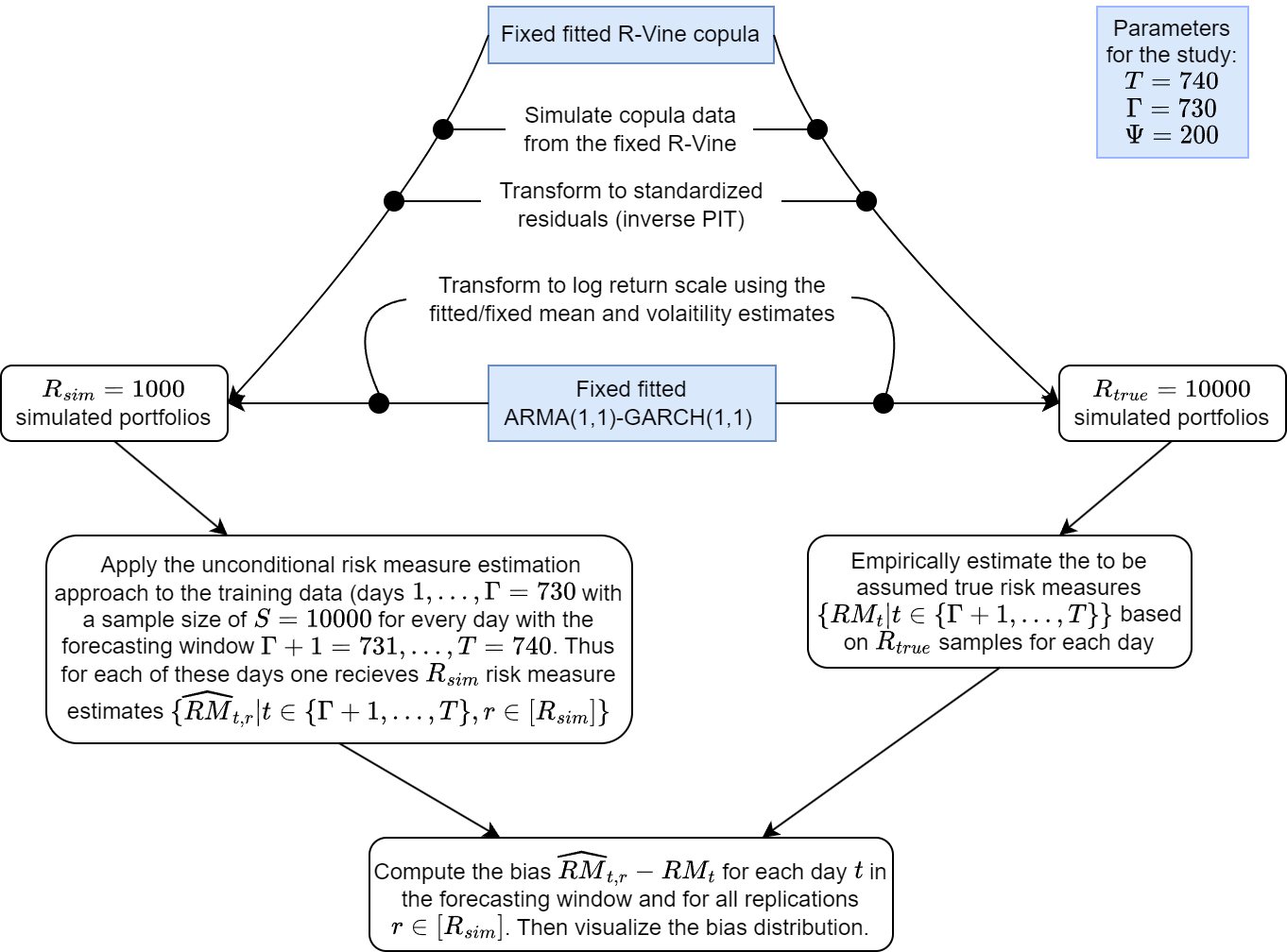}
\caption{An overview of the setup of the presented proof of concept simulation study.}
\label{fig:simstudyoverview}
\end{figure}

\blue{
More specifically we use as the ground truth for the time series the fitted values of respective fitted ARMA-GARCH models. This way the assumption that the marginals are learnable by an ARMA-GARCH model are satisfied by construction. As the ground truth for their dependence we take the corresponding fixed fitted R-vine copula. This mirrors our assumption that the flexible class of vine copulas is able to capture the multivariate dependence between the assets appropriately. Then we follow the simulation approach described in Figure \ref{fig:simstudyoverview}. }

\blue{
\noindent Without loss of generality we assume an equal weighting for all assets. We use a training horizon of $\Gamma = 730$ days and a validation horizon of $T-\Gamma = 740-730 = 10$ days without refits. All other parameters of our estimation approach were left with their standard values like $\Psi = 200$.\footnote{The detailed study with the code to follow along can be found at \url{https://github.com/EmanuelSommer/PortvineThesis/tree/master/analysis/poc_uncond_simulation}.}
}

\clearpage

\begin{figure}[hbt!]
    \centering
    \includegraphics[width = 0.9\textwidth]{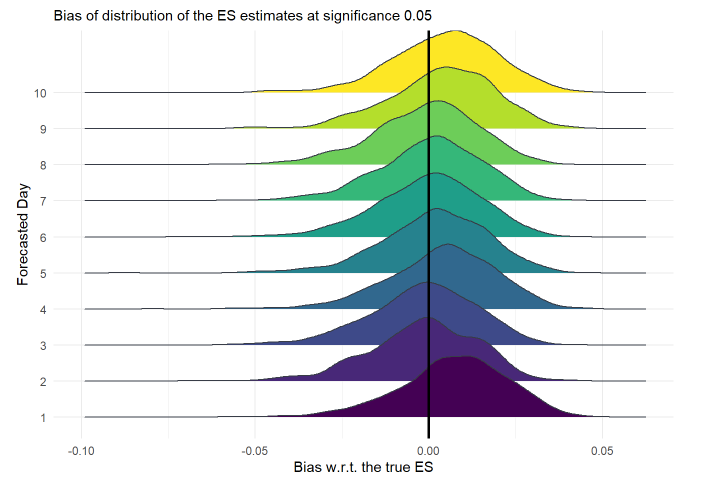}
    \caption{Bias of distribution of the simulation based $ES$ estimates at $\alpha = 0.05$ against the reference $ES$.}
    \label{fig:sim_study_005}
\end{figure}

\blue{
The resulting bias distributions of the ES estimates at the $\alpha = 0.05$ level by day in the validation set can be seen in Figure \ref{fig:sim_study_005}. We can observe that the estimation bias is almost always very close to 0 compared to the base variability of the log returns ($\hat\sigma = 0.083$). Furthermore we do not seem to systematically under or overestimate the ES. This is good for both the regulators which are averse against systematic overestimation and the financial institutions which are avers against systematic underestimation as it means elevated capital requirements. This is however still just a proof of concept simulation study with clear limitations like the estimation of the \textit{true} reference ES by rather small sample ($R_{true} = 10000$). All in all this small PoC study did not uncover any obvious problems in the stated estimation approach and suggests that a large scale simulation study is the next step to take.
}

\clearpage
\bibliography{mybibfile}

\end{document}